%% file: toi732.tex
\documentclass[twocolumn,twocolappendix]{aastex63}
\hypersetup{linkcolor=blue,citecolor=blue,filecolor=blue,urlcolor=blue}
\usepackage{amsmath}
\newcommand{\gaia}{\emph{Gaia}}

\newcommand{\tess}[1]{\emph{TESS}#1}

\newcommand{\teff}[1]{$T_{\text{eff}}$#1}
\newcommand{\prot}[1]{$P_{\text{rot}}$#1}
\newcommand{\logg}[1]{$\log{g}$#1}
\newcommand{\vsini}[1]{$v\sin{i}$#1}
\newcommand{\mps}[1]{m s$^{-1}$#1}
\newcommand{\teq}[1]{$T_{\text{eq}}$#1}
\newcommand{\Rearth}[1]{R$_{\oplus}$#1}
\newcommand{\Rsun}[1]{R$_{\odot}$#1}
\newcommand{\Mearth}[1]{M$_{\oplus}$#1}
\newcommand{\Msun}[1]{M$_{\odot}$#1}
\defcitealias{cloutier20}{CM20}

\usepackage{dashbox,framed,color,ocg-p}
\shortauthors{Cloutier et al.}
\shorttitle{Two planets spanning the radius valley around LTT 3780}

\begin{document}
\title{A pair of TESS planets spanning the radius valley around the 
  nearby mid-M dwarf LTT 3780}

\suppressAffiliations
\input{allauthors}

\correspondingauthor{Ryan Cloutier}
\email{ryan.cloutier@cfa.harvard.edu}

\begin{abstract}
  We present the confirmation of two new planets
  transiting the nearby mid-M dwarf LTT 3780 (TIC 36724087, TOI-732, $V=13.07$,
  $K_s=8.204$, $R_s$=0.374 \Rsun{,} $M_s$=0.401 \Msun{,} d=22 pc).
  The two planet candidates are identified in a single \tess{}
  sector and are validated with reconnaissance spectroscopy, ground-based
  photometric follow-up, and high-resolution
  imaging. With measured orbital periods of $P_b=0.77$ days,
  $P_c=12.25$
  days and sizes $r_{p,b}=1.33\pm 0.07$ \Rearth{,} $r_{p,c}=2.30\pm 0.16$
  \Rearth{,} the two planets span the radius valley in period-radius space
  around low mass stars thus
  making the system a laboratory to test competing theories of the
  emergence of the radius valley in that stellar mass regime. By combining 63
  precise radial-velocity measurements from HARPS and HARPS-N,
  we measure planet masses of $m_{p,b}=2.62^{+0.48}_{-0.46}$ \Mearth{} and
  $m_{p,c}=8.6^{+1.6}_{-1.3}$
  \Mearth{,} which indicates that LTT 3780b has a bulk composition consistent
  with being Earth-like, while LTT 3780c likely hosts an extended H/He envelope.
  We show that the recovered planetary masses are consistent with predictions
  from both photoevaporation and from core-powered mass loss models. 
  %The physical and orbital planet parameters, combined with the brightness
  %and small size of LTT 3780, render both of the known LTT 3780 planets
  %as accessible targets for atmospheric characterization of planets within
  %the same planetary system and spanning the radius valley.
  The brightness and small size of LTT 3780, along with the measured
  planetary parameters, render LTT 3780b and c
  as accessible targets for atmospheric characterization of planets within
  the same planetary system and spanning the radius valley.
\end{abstract}

\section{Introduction}
Since the commencement of its prime mission in July 2018, NASA's Transiting
Exoplanet Survey Satellite (\tess{;} \citealt{ricker15}) has unveiled many of
the closest transiting exoplanetary systems to our solar system. The proximity
of many of these systems make their planets ideal targets for the detailed
characterization of their bulk compositions and atmospheric properties.
Systems of multiple transiting planets are of particular interest as they
afford the unique opportunity for direct comparative planetology, having formed
within the same protoplanetary disk and evolved around the same host star.

The occurrence rate of close-in planets features a dearth of planets between
$1.7-2.0$ R$_{\oplus}$ around Sun-like stars and between $1.5-1.7$ around
low mass stars \citep{fulton17,mayo18,cloutier20,hardegree20}.
The so-called radius valley is likely a result 
of the existence of an orbital separation-dependent transition between primarily rocky
planets and non-rocky planets that host extended H/He envelopes.
A number of physical processes have been proposed to explain the existence of
this rocky/non-rocky transition, including photoevaporation, wherein XUV
heating from the host star drives thermal atmospheric escape preferentially
on smaller, low surface gravity planets during the first 100 Myrs
\citep{owen13,jin14,lopez14,chen16,owen17,jin18,lopez18,wu19}. Alternatively,
the core-powered mass loss mechanism, wherein the dissipation of the planetary
core's primordial energy from formation drives atmospheric mass loss over Gyr
timescales \citep{ginzburg18,gupta19,gupta20}. Rather than resulting from
the dissipation of primordial planetary atmospheres, the radius
valley may instead arise from the superposition of rocky and non-rocky planet
populations, with the former forming in a gas-poor environment after the
dissipation of the gaseous protoplanetary disk  \citep{lee14,lee16,lopez18}.

Each of the aforementioned
mechanisms make explicit predictions for the location of the rocky/non-rocky
transition in the orbital period-radius space. Measurements of
planetary bulk compositions in systems of multiple planets that span the radius
valley therefore offer an opportunities to resolve the precise location of the
rocky/non-rocky transition \citep{owen20} and distinguish between the model
predictions. Precise planetary bulk composition measurements for systems around
a range of host stellar masses will enable the dependence of the radius valley
on stellar mass to be
resolved and consequently used to test competing models of the emergence of
the radius valley \citep[][hereafter \citetalias{cloutier20}]{cloutier20}.

Here we present the discovery and confirmation of the two-planet system around
the nearby (d=22 pc) mid-M dwarf LTT 3780 from the \tess{} mission.
The planets LTT 3780b and c span the rocky/non-rocky transition such that the
characterization of their bulk compositions can be used to constrain emergence
models of the radius valley by marginalizing over unknown system parameters
such as the star's XUV luminosity history.
The brightness of LTT 3780 ($K_s=8.204$) and the
architecture of its planetary system also make it an attractive target for the
atmospheric characterization of multiple planets within the same planetary
system. In Sect.~\ref{sect:star} we present the properties of LTT 3780.
In Sect.~\ref{sect:observations} we present the \tess{} light curve along with
our suite of follow-up observations, including reconnaissance spectroscopy,
ground-based photometry, high-resolution imaging, and precise radial-velocity
measurements. In Sect.~\ref{sect:analysis} we present our two independent
analyses of our data, to ensure the robustness of our 
results, before concluding with a discussion and summary of our results in
Sects.~\ref{sect:discussion} and~\ref{sect:summary}.

\section{Stellar Characterization} \label{sect:star}
LTT 3780 (LP 729-54, TIC 36724087, TOI-732) is a mid-M dwarf at a distance of
22 pc \citep{gaia18,lindegren18}. Astrometry, photometry, and the LTT 3780
stellar parameters are reported in \autoref{tab:star}. The stellar
\teff{} $=3331\pm 157$ K is taken from the \tess{} Input Catalog
\citep[TIC v8;][]{stassun19} and is consistent with the value derived from the
Stefan-Boltzmann equation ($3343\pm 150$ K). The stellar metallicity is
weakly constrained by its SED and \texttt{MIST} isochrones \citep{dotter16}.
The LTT 3780 mass and radius
are derived from the stellar parallax and $K_s$-band magnitude, used to
compute the absolute $K_s$-band magnitude $M_{K_s}$, and the empirically-derived
M dwarf mass-luminosity and radius-luminosity relations from 
\cite{benedict16} and \cite{mann15} respectively.
LTT 3780's surface gravity is computed from its mass and radius.
No photometric rotation period
is apparent in either the \tess{} or ground-based photometry. However, the
low value of $\log{R_{\text{HK}}'} = -5.59$ is indicative of a chromospherically
inactive star with likely a long rotation period
\citep[estimated \prot{} $=104\pm 15$ days;][]{astudillodefru17b}.

\input{ltt3780table}

LTT 3780 is the primary component of a visual binary system with an angular
separation of $16.1''$ from the \gaia{} DR2 positions
\citep{gaia18,lindegren18}.
The binary was previously identified to be co-moving from measures of each
stellar component's proper motion and spectroscopic distance
\citep{luyten79,scholz05}. The common parallaxes and proper motions of LTT 3780
(alias LP 729-54) and its stellar companion LP 729-55 (TIC 36724086) were
verified in \gaia{} DR2. Their angular separation of $16.1''$
implies a projected physical separation of 354 AU.
The fainter companion star has $K_s=10.223\pm 0.021$ (i.e.
$\Delta K_s=2.019$ mag) which corresponds to a mass and radius of
$0.136\pm 0.004$ \Msun{} and $0.173\pm 0.005$ \Rsun{.} %Based on each star's
%$K_s$-band magnitude ($K_{s,\text{LTT 3780}}=8.204, K_{s,\text{LP 729-55}}=10.223$),
%we derive a stellar mass ratio of $q=0.340\pm 0.014$ using the $K_s$-band
%mass-luminosity relation \citep{benedict16}.
Given the stellar mass ratio of $q=0.340\pm 0.014$, the orbital period of the
stellar binary at their projected physical separation
is about 9100 years. Assuming a circular orbit, this corresponds to a
negligible maximum radial velocity (RV) variation of $\lesssim 15$ cm s$^{-1}$
over the timescale of our RV observations presented in Sect.~\ref{sect:rvobs}.
We also calculated the secular acceleration of the binary system
given its large proper motion (\autoref{tab:star}) to be $<10$ cm s$^{-1}$
year$^{-1}$. This RV variation is also well below the noise limit of our
observations over our RV baseline.

The LTT 3780 planetary system may be an interesting test case of 
planet formation models in a binary systems. Although, the large physical separation
of the stellar components likely resulted in isolated planet formation around
LTT 3780.

\section{Observations} \label{sect:observations}
\subsection{TESS photometry} \label{sect:tessphot}
LTT 3780 was observed in \tess{} sector nine (i.e. orbits 25 and 26)
for 27.26 days from UT February 28 to March
26, 2019 with CCD 1 on Camera 1. As a member of the Cool Dwarf target list
\citep{muirhead18}, LTT 3780 was included in the TIC and in
the Candidate Target List \citep[CTL;][]{stassun17} such
that its light curve was sampled at 2-minute cadence.
These data were processed by the NASA Ames
Science Processing Operations Center \citep[SPOC;][]{jenkins16}.
The resulting Presearch Data Conditioning Simple Aperture Photometry
\citep[PDCSAP;][]{smith12,stumpe12,stumpe14} light curve of LTT 3780 was
corrected for dilution by known contaminating sources within the photometric
aperture with a dilution factor of 0.713. According to the sector nine data
release notes\footnote{\url{https://archive.stsci.edu/missions/tess/doc/tess_drn/tess_sector_09_drn11_v04.pdf}}, the level of scattered light from
the Earth in Camera 1 CCD 1 at the start of each orbit was high and resulted in
no photometry or centroid positions being calculated during the first
1.22 days of orbit 25 nor in the first 1.12 days of orbit 26. Data collection
was also paused for 1.18 days for data downloading close to the spacecraft's
time of perigee passage. Overall, a total of 24.08 days of science data
collection was performed in \tess{} sector nine.

\begin{figure}
  \centering
  \includegraphics[width=0.99\hsize]{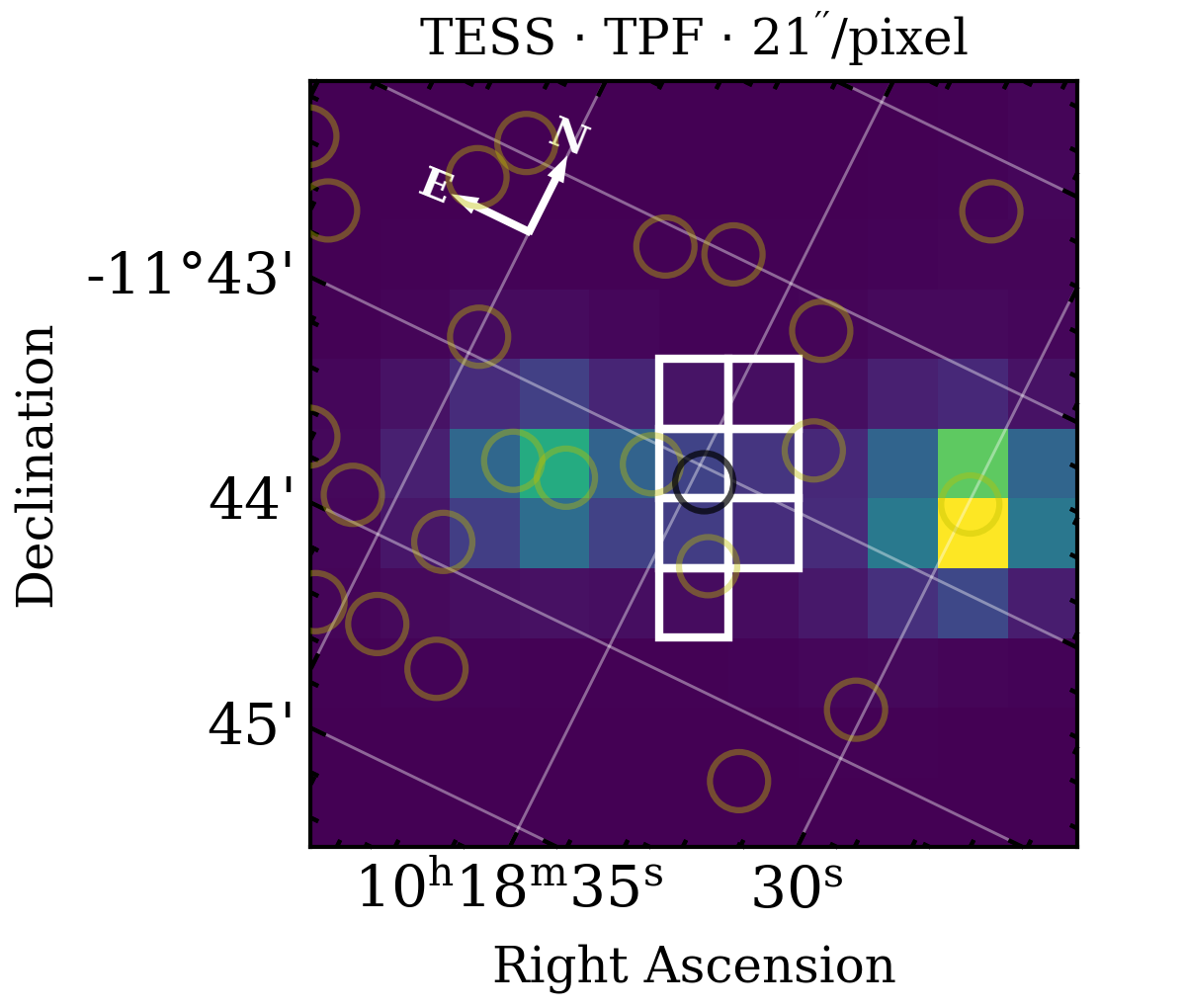}
  \includegraphics[width=0.99\hsize]{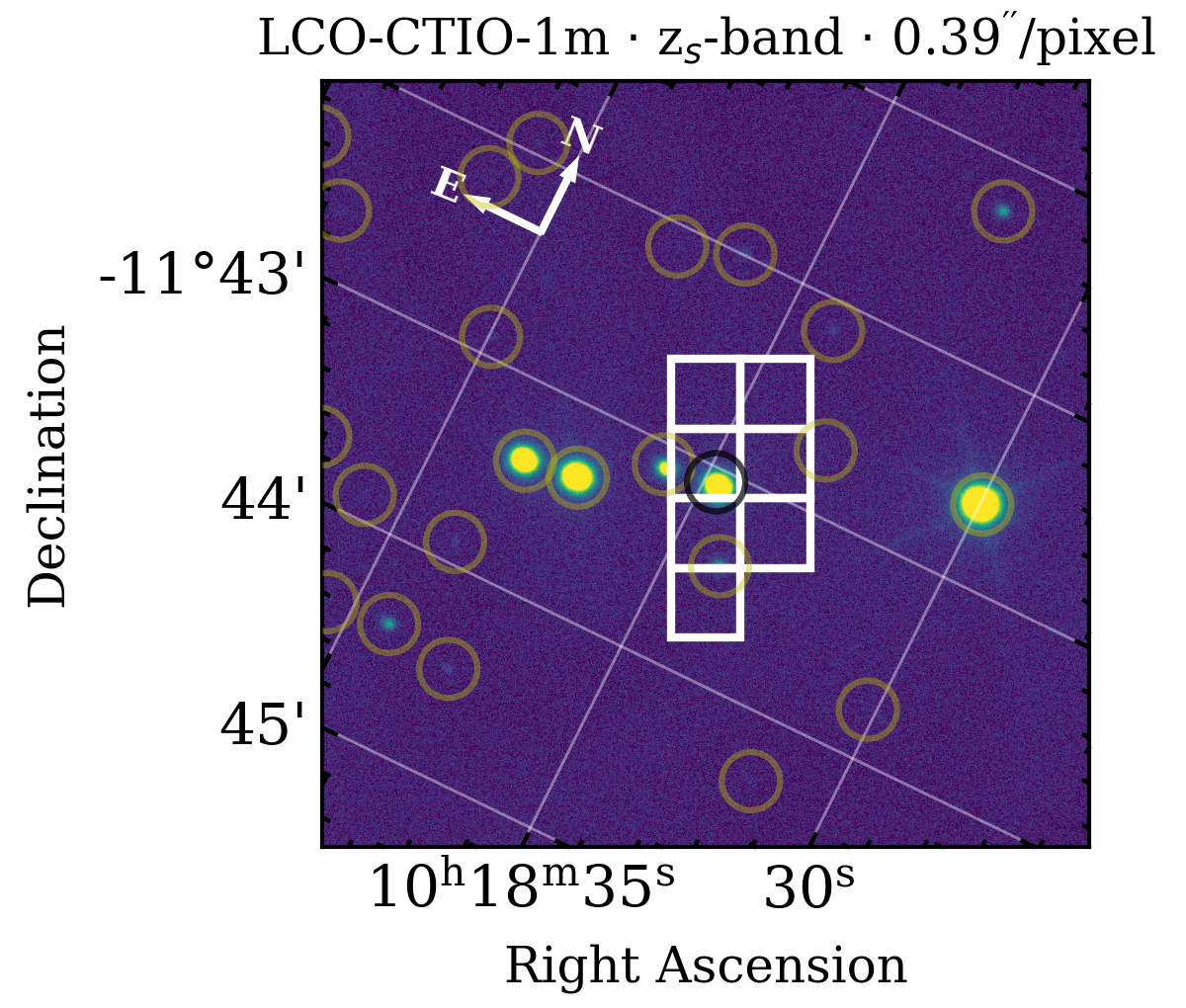}
  \caption{\emph{Upper panel}: an example \tess{} target pixel file image of
    LTT 3780 and the surrounding field. The \tess{} pixel scale is $21''$.
    The position of LTT 3780 in \gaia{}
    DR2 is circled in black while the remaining \gaia{} sources out to $2.5'$
    are circled in yellow. The pixels highlighted in white demarcate the
    \tess{} photometric aperture used to derive the PDCSAP light curve of LTT
    3780. \emph{Lower panel}: an example $z_s$ image of the same field taken
    with the LCOGT 1m telescope at CTIO with a much finer pixel scale
    of $0.39''$ pixel$^{-1}$ enabling LTT 3780 and nearby sources to be
    spatially resolved.}
  \label{fig:stamps}
\end{figure}

A sample image of the field surrounding LTT 3780 from the \tess{} target pixel
files is shown in \autoref{fig:stamps}. The \tess{} photometric
aperture used to produce the PDCSAP light curve was selected to maximize
photometric signal-to-noise \citep{bryson10} and is overlaid in
\autoref{fig:stamps}. 
Blending in the \tess{} photometry by nearby sources is unsurprising given
the large ($21''$) \tess{} pixels and the $1'$ FWHM of its point spread
function, coupled with the large number density of
37 sources within $2.5'$ \citep{gaia18,lindegren18}.
In \autoref{fig:stamps}, the low-resolution \tess{}
image is compared with an example ground-based image taken with the 1m telescope
at the Cerro Tololo Inter-American Observatory (CTIO) location of the Las
Cumbres Observatory Global Telescope network (LCOGT). The LCOGT $z_s$-band
image features a pixel scale of $0.39''$ which is equivalent to a spatial
resolution that is 54 times higher than in the \tess{} images. The
LCOGT image clearly depicts the
position of LTT 3780 within the \tess{} aperture and the positions of 24 nearby
sources from the \gaia{} DR2. The relative positions of the neighboring sources
to the \tess{} photometric aperture reveals how the aperture was optimized to
minimize contamination by the nearby bright sources including the binary
companion star LP 729-55 at $16.1''$ east of LTT 3780's position.

In the subsequent transit search conducted by the SPOC using the
Transiting Planet Search (TPS) Pipeline Module \citep{jenkins02,jenkins10},
two transiting
planet candidate signals were flagged and subsequently passed a set of internal
data validation tests \citep{twicken18,li19}. The planet candidates TOI-732.01
and 02 had reported periods of 0.768 days and 12.254 days,
corresponding to 28 and 2 observed transits respectively. However,
focusing solely on \tess{} measurements wherein the
quality flag \texttt{QUALITY} equals zero, indicating the reliability of those
measurements, the second transit of TOI-732.02 is only partially resolved as
its ingress is largely contaminated. Although the SPOC does not make an
identical cut based on the \texttt{QUALITY} flag, the SPOC-reported orbital
period of TOI-732.02 is found to be underestimated by about three minutes as we
will learn from our follow-up transit light curve
analysis (Sect.~\ref{sect:sg1}).

The initially reported depth for each planet
candidate was $1253\pm 106$ and $3417\pm 283$ ppm corresponding to preliminary
planetary radii of $1.44\pm 0.07$ and $2.38\pm 0.12$ R$_{\oplus}$ using the
stellar radius reported in \autoref{tab:star}. Note that
these planet parameters are preliminary and will be refined in our analysis
of the \tess{} light curve in Sect.~\ref{sect:tessmcmc}.

\subsection{Reconnaissance spectroscopy}
\subsubsection{TRES spectroscopy}
We obtained a single reconnaissance spectrum of LTT 3780 with the Tillinghast
Reflector \'Echelle Spectrograph (TRES), mounted on the
1.5m Tillinghast Reflector telescope at Fred L. Whipple Observatory (FLWO) on
Mount Hopkins, AZ on UT January 30, 2020. TRES is a fiber-fed,
$R=44,000$ optical \'echelle spectrograph (310-910 nm)
whose typical limiting RV precision on slowly rotating M dwarfs of 50 \mps{} is 
insufficient to measure the masses of the LTT 3780 planet candidates. We
obtained the spectrum to assess the star's level of
chromospheric activity, to potentially measure rotational broadening, and to
search for a double-lined spectrum that could indicate the
presence of a close-in stellar companion to LTT 3780.
We median-combined three 600 second exposures that were 
wavelength calibrated using a ThAr lamp exposure. The resulting
signal-to-noise (S/N) per resolution element at 715 nm
was 16. We then cross-correlated the spectrum order-by-order with an empirical
template spectrum of Barnard's star.

The reduced data revealed a single-lined spectrum. We see $H\alpha$ in
absorption and do not resolve any rotational broadening. With these data we
place an upper limit on \vsini{} at half the spectral resolution of TRES;
\vsini{} $\leq 3.4$ km s$^{-1}$. Note that this value will be refined in
Sect.~\ref{sect:rvobs} with our high resolution spectra from HARPS.
The lack of $H\alpha$ in emission and lack
of any significant stellar rotation, combined with the low level of stellar
photometric variability in the \tess{} light curve and the absence of flares,
emphasizes the low levels of magnetic activity produced by LTT 3780.
This fact will have important implications for the precise RV
characterization of the TOI-732 planetary system and for
future atmospheric characterization efforts in which atmospheric feature
detections may be degenerate with signatures from magnetically active regions
if not properly modeled in transmission spectra \citep{rackham18}.

\subsection{Ground-based transit photometry} \label{sect:sg1}
\tess{'}s large pixels ($21''$) result in significant blending of the LTT 3780
light curve with nearby sources, including with its visual binary companion at
$16.1''$ to the east
(with a \tess{} magnitude difference $\Delta T=2.42$,
see \autoref{fig:stamps}).
We obtained seeing-limited photometric follow-up observations of the LTT 3780
field close to the expected transit times of each planet candidate 
as part of the \tess{} Follow-up Observing Program (TFOP). The example image
from this follow-up campaign in \autoref{fig:stamps} reveals how individual
sources are resolved, which enabled the confirmation of the transit events
on-target, and the scrutiny of nearby sources for nearby eclipsing
binaries (EBs).
Follow-up efforts were scheduled using the \tess{} Transit Finder, which is a
customized version of the \texttt{Tapir} software package \citep{jensen13}. 
Unless otherwise noted, the photometric data were extracted and detrended using
the \texttt{AstroImageJ} software package \citep[AIJ;][]{collins17}. The
resulting light curves were detrended with any combination of time (i.e. a
linear trend), airmass, and total background counts as necessary in attempts to
flatten the out-of-transit portion of each light curve. Furthermore, the
differential light curves were derived using an optimal photometric aperture and
a set of comparison stars chosen by the observer.

\begin{figure}
  \centering
  \includegraphics[width=\hsize]{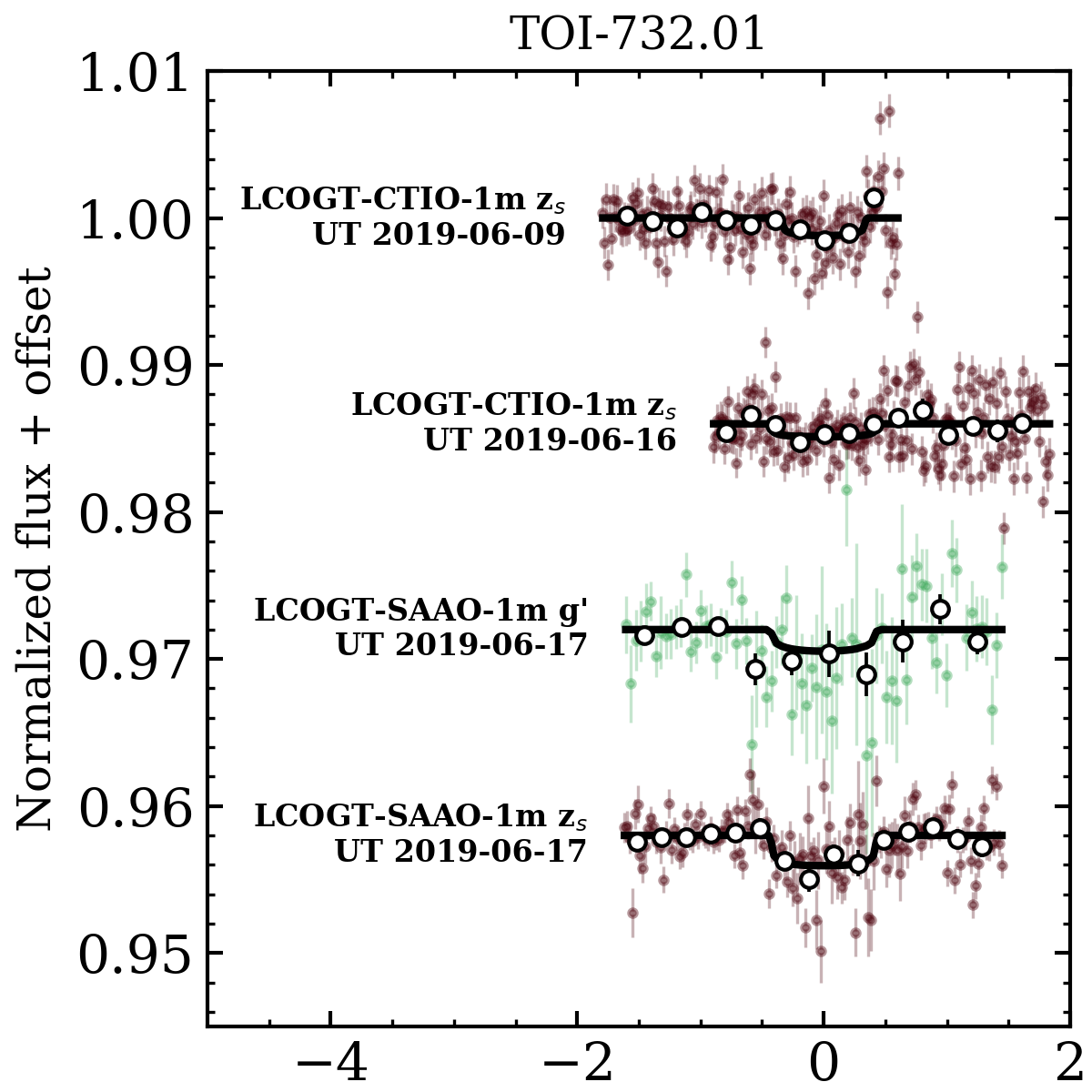}
  \includegraphics[width=\hsize]{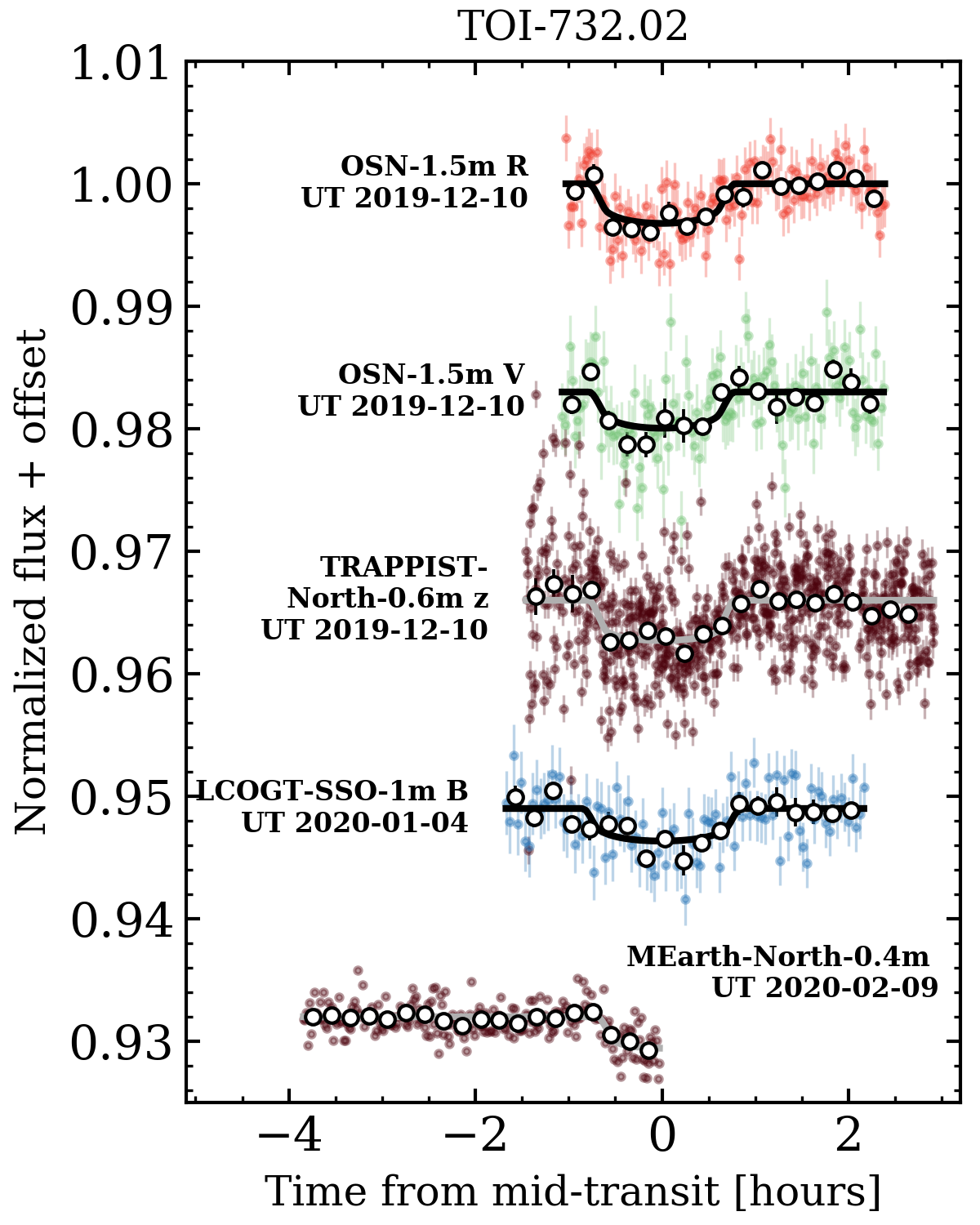}
  \caption{Ground-based transit light curves of TOI-732.01 (\emph{upper panel})
    and 02 (\emph{lower panel}) taken
    as part of TFOP. Solid curves depict the optimized transit model fit with
    all model parameters fixed other than the baseline flux,
    the mid-transit time, and the planet-to-star radius ratio.
    Annotated next to each light curve is the telescope facility, the passband,
    and the UT observation date.}
  \label{fig:sg1}
\end{figure}

Numerous ground-based facilities conducted photometric follow-up of the TOI-732
system. Their respective data acquisition and reduction strategies are
described in the following sections while their detrended light curves are
plotted in \autoref{fig:sg1}.
Differences in the instrumental setups and nightly observing conditions
produce varying levels of photometric precision among the light curves.
Each detrended light curve,
available through TFOP, is fit with a \cite{mandel02} transit model that we 
calculate using the \texttt{batman} software package \citep{kreidberg15}.
The shallow transit depths of both planet candidates produce low
S/N transit light curves that may only marginally
improve the measurement precision on most model parameters compared to
the values measured from the \tess{} light curve with the exception being the
planets' orbital periods when all light curves are fit simultaneously. As such,
we fix the orbital periods and impact parameters in the individual light curve
fits to the values obtained from the SPOC Data
Validation module ($P_b=0.76842$ days, $P_c=12.25422$ days, $b_b=0.69$,
$b_c=0.35$). We also derive the scaled semimajor axes using the
stellar parameters given in \autoref{tab:star} ($a_b/R_s=6.96, a_c/R_s=44.09$).
Each planet's orbit is also fixed to circular and the quadratic limb darkening
parameters in the corresponding passband are interpolated from the
\cite{claret11} tables using the \texttt{EXOFAST} software \citep{eastman13}
given LTT 3780's \teff{,} \logg{,} and [Fe/H]. We fit the
following parameters via non-linear least squares optimization using
\texttt{scipy.curve\_fit}: the baseline flux  
$f_0$, the time of mid-transit
$T_0$, and the planet-to-star radius ratio $r_p/R_s$. Measuring $T_0$ relative
to the expected transit time is used to refine the
planet's orbital ephemeris while $r_p/R_s$ measurements in each passband are
required to investigate transit depth chromaticity as a chromatically varying
transit depth could be indicative of a blended EB.

\subsubsection{LCOGT photometry}
We used three observatories as part of the Las Cumbres Observatory Global
Telescope network \citep[LCOGT;][]{brown13} to follow-up transits of
both TOI-732.01 and 02. Each 1m telescope is equipped with a
$4096\times 4096$ LCOGT SINISTRO camera whose pixel scale is 389 mas
pixel$^{-1}$, resulting in a $26' \times 26'$ field-of-view (FOV).
We calibrated all image sequences using the standard LCOGT \texttt{BANZAI}
pipeline \citep{mccully18}. An example of one such image from the LCOGT was
shown in \autoref{fig:stamps}.

We observed three full transits of TOI-732.01 between UT June 9-17, 2019
from various LCOGT observatories. These data include two $z_s$-band light curves
taken at the LCOGT-Cerro Tololo Inter-American Observatory (CTIO) on UT June 9
and 16 2019, and a third transit light curve obtained on UT June 17, 2019 in the
$z_s$ and $g'$-bands by the LCOGT-South African Astronomical Observatory (SAAO).
These four light curves are shown in \autoref{fig:sg1}.
We searched for transit-like events from nearby EBs (NEB)
around 37 sources identified by \gaia{} DR2 to be within $2.5'$.
The field was consequently cleared of NEBs down to $\Delta z_s=7.686$
as no transit-like signals were detected on any off-target source.
All three expected transit events were shown to occur on-target
and arrived within 4 minutes of their expected transit times.

We observed one full transit of TOI-732.02 on UT January 4, 2020 with the
LCOGT-Siding Springs Observatory (SSO) in the $B$-band. The light curve is
included in \autoref{fig:sg1}. Similarly to our TOI-732.01 transit analysis,
the field was cleared of NEBs during the TOI-732.02 transit window. The
expected transit event was shown to occur on-target with a transit depth
of 2.4 parts per thousand (ppt). However, the transit 
arrived 60 minutes early indicating that the preliminary orbital period of
$P_c=12.254$ days, derived from the \tess{} light curve alone, is slightly
underestimated if the period is constant.
%Recall that the ingress of the
%second TOI-732.02 transit in the \tess{} light curve was missed such that the
%misestimate of its preliminary orbital period is unsurprising.
The orbital period of LTT 3780c will be refined in our global analysis in
Sect.~\ref{sect:analysis}, which will include the ground-based light curves.

\subsubsection{OSN photometry}
We observed one additional transit of TOI-732.02 on UT December 10, 2019 with
the Observatorio de Sierra Nevada (OSN) 1.5m telescope near Granada, Spain.
The OSN 1.5m telescope is equipped with an Andor ikon-L $2048\times 2048$ CCD
camera whose pixel scale is 232 mas pixel$^{-1}$, resulting in a
$7.9' \times 7.9'$ FOV.
We observed the full transit simultaneously in both the $V$ and $R$-bands to
check for chromaticity. Similarly to the LCOGT-SSO transit observation of
TOI-732.02, the expected transit event arrived 60 minutes early.
The measured transit depths of 2.9 ppt and 3.2 ppt in the $V$ and
$R$-bands respectively are consistent with each other and with
the LCO-SSO $B$-band transit at $1\sigma$.
TOI-732.02 therefore does not show any strong
chromaticity. The two transit light curves are included in \autoref{fig:sg1}.

\subsubsection{TRAPPIST-North photometry}
The UT December 10, 2019 transit of TOI-732.02 observed by OSN was also observed
by the 60cm TRAnsiting Planets and PlanetesImals Small Telescope-North
(TRAPPIST-North) located at the Ouka\"imden Observatory in Morocco
\citep{jehin11,gillon13c,barkaoui19}.
TRAPPIST-North employs a $2048\times 2048$ pixel Andor IKONL BEX2 DD camera
with a pixel scale of 600 mas pixel$^{-1}$ resulting in a $20.5' \times 20.5'$
FOV. The photometry was analyzed using custom software for TRAPPIST-North.
We observed the full transit in the $z$-band, thus contributing to the
four transit light curves of TOI-732.02 from TFOP in the $B$, $V$, $R$, and
$z$-bands. The measured transit depth in the $z$-band is 3.2 ppt, which
is consistent with the measured transit depths in the aforementioned passbands
thus confirming that no strong chromaticity is detected. The TRAPPIST-North
light curve is included in \autoref{fig:sg1}.

\subsubsection{MEarth-North photometry}
We observed a partial transit of TOI-732.02 on UT February 9, 2020 using
seven of eight telescopes from the
MEarth-North telescope array located at FLWO on Mount Hopkins, AZ. The
MEarth-North array consists of eight 40cm Ritchey-Chr\'etien telescopes, each
equipped with a $2048\times 2048$ pixel Apogee U42 camera. The 750 mas pixel
scale results in a $25.6' \times 25.6'$ FOV. The light
curve was obtained in the custom MEarth passband centered in the red optical and
is shown in \autoref{fig:sg1}. The
observations include a three hour out-of-transit baseline plus the transit
ingress and 37 minutes in-transit, equal to nearly half of the full transit
duration. The measured transit depth of 3.3 ppt is consistent with all
other TFOP transits again confirming the lack of transit depth chromaticity.

The collective photometric data from TFOP have verified the periodic nature
of the transits of TOI-732.01 and 02 and that both of these planet
candidates orbit the target star LTT 3780. We do not detect any significant
depth discrepancies, indicating that the transits are likely achromatic and thus
consistent with being planetary in origin. Furthermore, the early arrival of
the TOI-732.02 transits on December 10, 2019 and on January 4, 2020 allow
us to estimate the true orbital period of LTT 3780c, which shrinks from its
SPOC-reported value of 12.254 to 12.2519 days,
assuming a constant period. This refined period prior is used in our
up-coming analysis of the \tess{} light curve in Sect.~\ref{sect:tessmcmc}.

\subsection{High-resolution imaging}
Very nearby stars that are not detected in \gaia{} DR2, nor in any of the
seeing-limited image sequences, and that fall within the same $21''$ \tess{}
pixel as the target star, will result in photometric contamination that is
unaccounted for in the \tess{} light curve. This effect reduces the depth of
the observed transits and can produce a false positive transit signal from
another astrophysical source, such as a blended EB \citep{ciardi15}.
We used two independent sets of high-resolution follow-up imaging sequences 
to search for any such close-in sources as described in the following sections.

\subsubsection{SOAR speckle imaging}
\begin{figure}
  \centering
  \includegraphics[width=0.98\hsize]{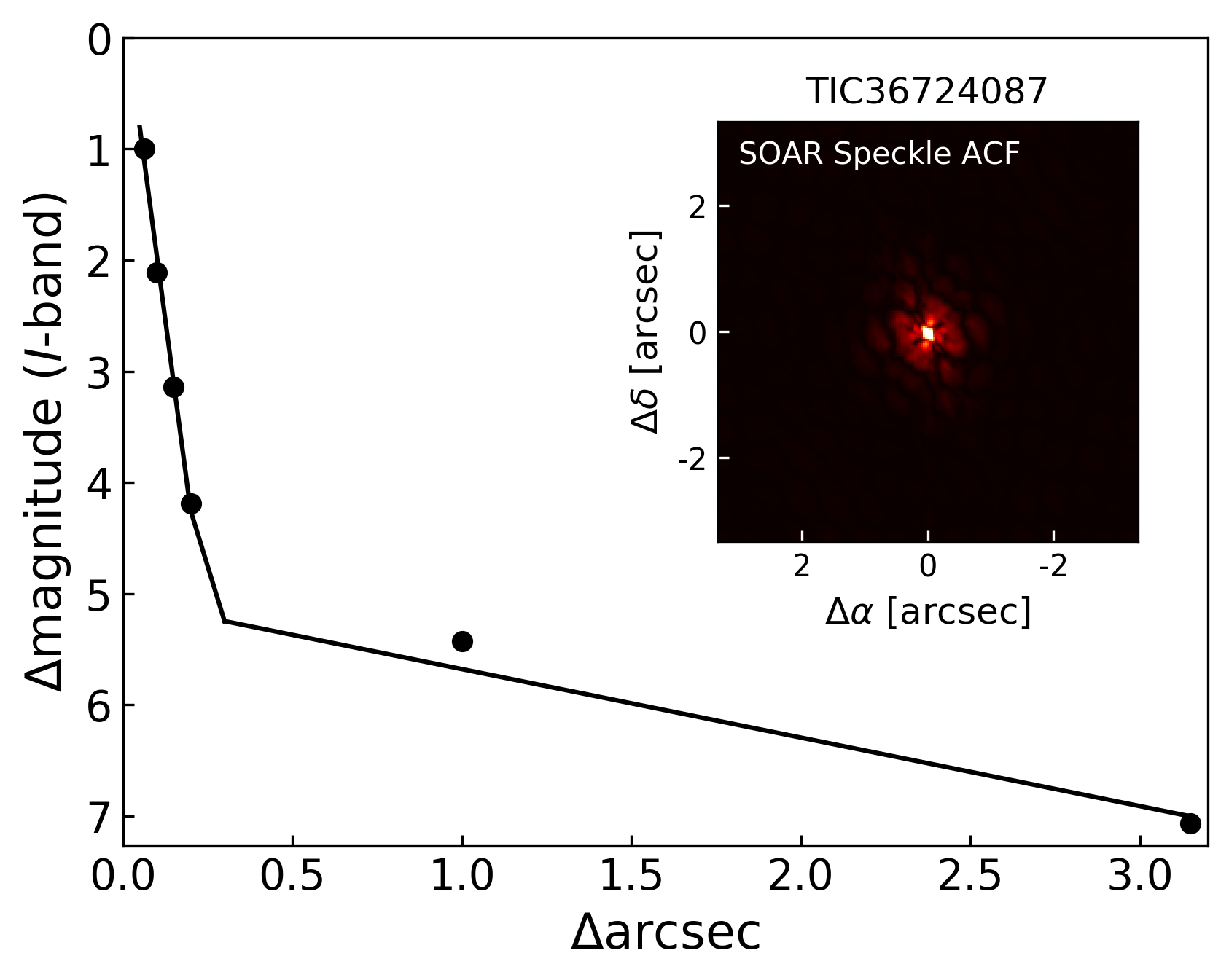}
  \includegraphics[width=0.98\hsize]{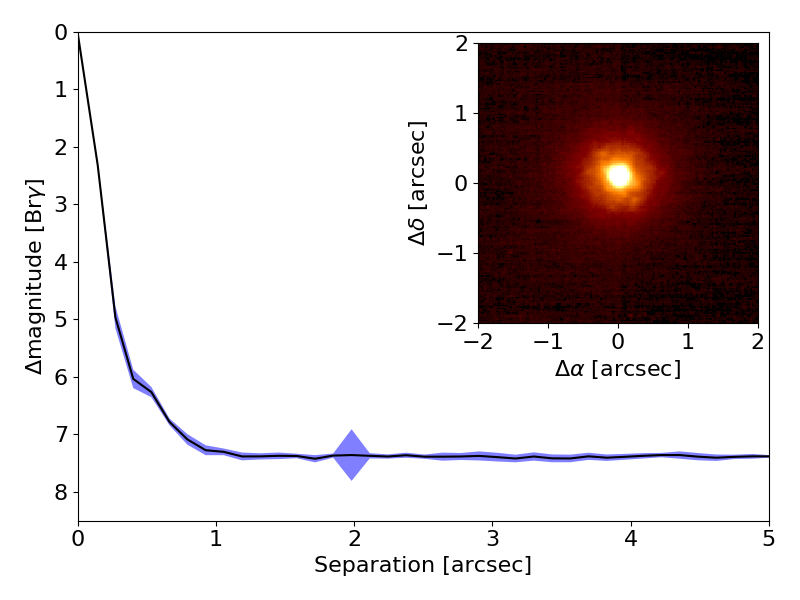}
  \caption{\emph{Upper panel}: $I$-band $5\sigma$ contrast curve from
    \emph{SOAR}
    speckle imaging of LTT 3780 (TIC 36724087). The inset depicts the
    corresponding speckle auto-correlation function. \emph{Lower panel}:
    Br$\gamma$ $5\sigma$ contrast curve from \emph{Gemini}/NIRI AO
    imaging. A few bad pixels persist at $2''$ from the target
    (\emph{blue diamond}), but these have a minimal effect on the contrast. 
    The inset depicts the central coadded image centered on LTT 3780.
    No visual companions  are detected in either dataset at $\geq 5\sigma$.}
  \label{fig:imaging}
\end{figure}

We obtained \emph{SOAR} speckle imaging \citep{tokovinin18} of LTT 3780 on
UT December 12, 2019 in the $I$-band, a visible bandpass similar to that of
\tess{.} Details of the observations from the \emph{SOAR} \tess{}
survey are provided in \cite{ziegler20}.  No bright nearby stars are detected
within $3''$ of LTT 3780 within the $5\sigma$ detection sensitivity of the
observations.
The resulting $5\sigma$ contrast curve is plotted in \autoref{fig:imaging}
along with the speckle auto-correlation function.

\subsubsection{NIRI AO imaging}
We obtained adaptive-optics (AO) images with \emph{Gemini}/NIRI \citep{hodapp03}
on UT November 25, 2019 in the Br$\gamma$ filter (2.17 $\mu$m).
We collected nine dithered
images with integration times of 2.2 seconds. We followed a standard data
reduction procedure including corrections for bad pixels, flat-fielding, sky
background subtraction, and image coaddition. No visual companions are
identified within $5''$ of LTT 3780 within the $5\sigma$ sensitivity
of the observations. These high quality data are sensitive to companions five
magnitudes fainter than the target at just 270 mas and 7.4 magnitudes fainter
at separations $\gtrsim 1''$. The $5\sigma$ contrast curve and the
coadded image centered on LTT 3780 are included in \autoref{fig:imaging}.

Due to the single-lined spectrum of LTT 3780, the verification of the expected
transit events on-target from ground-based photometry, and the lack of nearby
contaminating sources from high-resolution imaging,
we conclude that the planet candidates TOI-732.01 and 02 are verified planets.
We will refer to these planets as LTT 3780b and c for the remainder
of this study.

\subsection{Precise radial-velocities} \label{sect:rvobs}
\subsubsection{HARPS radial velocities}
We obtained 33 spectra of LTT 3780 with the High Accuracy
Radial velocity Planet Searcher \citep[HARPS;][]{mayor03} \'echelle spectrograph
mounted at the ESO 3.6m telescope at La Silla Observatory, Chile. The
HARPS optical spectrograph at $R=115,000$ is stabilized in pressure and
temperature, which enable it to achieve sub-\mps{} accuracy. The
observations were taken between UT June 21, 2019 and February 24, 2020 as
part of the ESO program 1102.C-0339. The exposure time was set to 2400 seconds,
which resulted in a median S/N over all orders of 26 and
a median measurement uncertainty of 1.31 \mps{} following the RV extraction
described below. Similarly to the TRES reconnaissance spectra at
$R=44,000$, LTT 3780 does not exhibit any rotational broadening in the HARPS
spectra. The corresponding upper limit on stellar rotation is \vsini{}
$\leq 1.3$ km s$^{-1}$. 

We extracted the HARPS RV measurements using the \texttt{TERRA} pipeline
\citep{anglada12}. \texttt{TERRA} employs a template-matching scheme that has
been shown to outperform the cross-correlation function (CCF) technique on
M dwarfs \citep{anglada12}.
M dwarfs are particularly well-suited to RV extraction via template-matching
because the line lists used to define the binary mask
for the CCF technique are incomplete. The resulting CCF template is often a poor
match for cool M dwarfs.

\texttt{TERRA} constructs a master template spectrum by first shifting the
individual spectra to the barycentric frame using the barycentric corrections
calculated by the HARPS Data Reduction Software \citep[DRS;][]{lovis07}. 
We masked portions of the wavelength-calibrated spectra in which telluric
absorption exceeds 1\%. The spectra are then coadded to build a high S/N
spectral template. We computed the RV of each spectrum by least-squares
matching the individual spectrum to the master template. 
Throughout the extraction process, we only consider orders redward of order 18
(428-689 nm) such that the bluest orders at low S/N are ignored. Because the
master spectrum is derived from the observed spectra, template construction
does not require any additional assumptions about the stellar properties. Using
this method, we found that the median LTT 3780 RV measurement precision was
improved by a factor of two compared to the
standard CCF method utilized within the HARPS DRS. The resulting RV time
series is reported in \autoref{tab:rvs}.

%The RV measurements were computed via a maximum likelihood analysis between a
%master stellar template and the individual spectra using the
%\texttt{NAIRA} routines \citep{astudillodefru17a}.
%The master stellar template was constructed from the
%median of all the spectra following their shift to the stellar frame.
%Similarly, a telluric template was derived from the median of the spectra
%shifted to the Earth's frame. In each of these two steps we used the stellar RV
%derived by the HARPS Data Reduction Software \citep[DRS;][]{lovis07} through a
%cross-correlation function (CCF). The correction for the barycentric motion of
%the Earth was also computed using the HARPS DRS. The resulting stellar template
%was Doppler shifted over a 40 km s$^{-1}$ wide window and centered on the
%average of the DRS RVs. The telluric template was used to mask the regions of
%the spectrum that were significantly contaminated by telluric features. For
%each RV step the value of the likelihood function was evaluated and the RV
%of the spectrum was calculated using maximum likelihood statistics. This
%process was repeated for the full set of HARPS spectra. The resulting RV time
%series is reported in \autoref{tab:rvs}.

\input{rvtable}

\subsubsection{HARPS-N radial velocities}
We obtained 30 spectra of LTT 3780 with the HARPS-N optical
\'echelle spectrograph at the TNG on La Palma in the Canary Islands. The
observations
were taken as part of the HARPS-N Collaboration Guaranteed Time Observations
program between UT December 14, 2019 and March 15, 2020. The exposure time
was set to 1800 seconds, which resulted in a median S/N over all orders of 20
and a median measurement uncertainty of 1.43 \mps{.}

Identically to the HARPS RVs, we extracted
the HARPS-N RVs using the \texttt{TERRA} template-matching
algorithm. The resulting RV time series is included in \autoref{tab:rvs}.

%The RVs were extracted using the \texttt{TERRA} pipeline \citep{anglada12}.
%\texttt{TERRA} employs a template-matching scheme that has been shown to
%outperform the cross-correlation function (CCF) technique for RV extraction on
%M dwarfs. The median RV measurement precision obtained from the \texttt{TERRA}
%extraction was improved by a factor of $\sim 2$ compared to the standard CCF
%method utilized within the HARPS-N DRS \citep{lovis07}. The resulting RV time
%series is included in \autoref{tab:rvs}.

\section{Data Analysis \& Results} \label{sect:analysis}
Here we conduct two independent analyses of our data
to test the robustness of the recovered
planetary parameters. In our fiducial analysis (Sects.~\ref{sect:tessmcmc}
and~\ref{sect:rvs}), the \tess{} light curve is modeled separately with the
resulting planet parameters being used as priors in the subsequent RV analysis.
In Sect.~\ref{sect:exofast} we describe an alternative, global analysis using
the \texttt{EXOFASTv2} software \citep{eastman19}.

\subsection{TESS transit analysis} \label{sect:tessmcmc}
\begin{figure*}
  \centering
  \includegraphics[width=0.99\hsize]{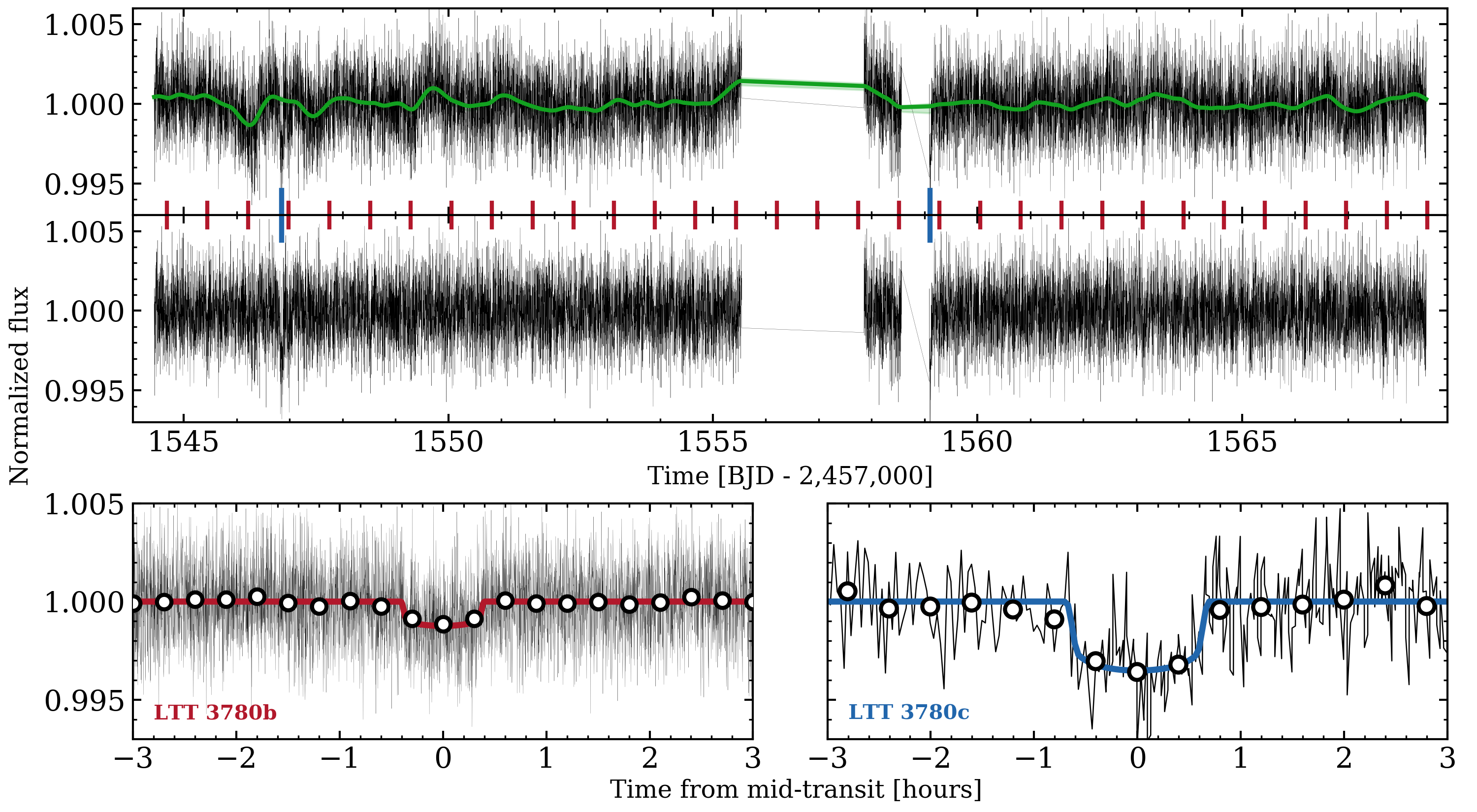}
  \caption{\emph{Upper panel}: the \tess{} PDCSAP light curve of LTT 3780
    (\emph{black curve}) along with the mean GP detrending model
    (\emph{green curve}) and its $3\sigma$ confidence interval in the
    surrounding shaded region which is narrow and hence difficult to discern.
    The \emph{vertical red} and \emph{blue ticks} along the x-axis highlight
    the mid-transit times of the planets
    LTT 3780b and c respectively. \emph{Middle panel}: the detrended \tess{}
    light curve. \emph{Lower panels}: phase-folded light curves of LTT 3780b
    (\emph{left}) and c (\emph{right}) along with their best-fit transit
    models. \emph{White markers} depict the temporally-binned phase-folded
    light curves to help visualize the transit events.}
  \label{fig:tess}
\end{figure*}

We begin by analyzing the \tess{} PDCSAP light curve wherein the planet
candidates TOI-732.01 and 02 were initially detected.
The majority of apparent
signals from non-random noise sources in the light curve have already been
removed by the SPOC processing.
However, low frequency and small amplitude signals that do not
resemble planetary transits are seen to persist in the PDCSAP light curve
shown in \autoref{fig:tess}. The
nature of these signals as residual systematics or photometric stellar
variability is unclear so we proceed with modeling the aforementioned
noise signals as an untrained semi-parametric Gaussian process (GP) regression
model, simultaneously with the two transiting planet candidates using the
\texttt{exoplanet} software package \citep{foremanmackey19}. \texttt{exoplanet}
computes analytical transit models using the \texttt{STARRY} package
\citep{luger19} and uses the \texttt{celerite} package
\citep{foremanmackey17} to evaluate the
marginalized likelihood under a GP model. In this analysis, the covariance
kernel takes the form of a stochastically-driven, damped simple harmonic
oscillator (SHO) whose Fourier transform is known as the power spectral density
(PSD) and is given by

\begin{equation}
  S(\omega) = \sqrt{\frac{2}{\pi}}
  \frac{S_0\,\omega_0^4}{(\omega^2-{\omega_0}^2)^2 + {\omega_0}^2\,\omega^2/Q^2}.
\end{equation}

\noindent The PSD of the SHO is parameterized by the frequency of the undamped
oscillator $\omega_0$, $S_0$, which is proportional to the power at the
frequency $\omega_0$, and the quality factor $Q$, which is fixed to $\sqrt{0.5}$.
We selected this covariance kernel and parameterization because
working in Fourier space is much more computationally efficient for large
datasets, such as our \tess{} light curve ($N=15,210$), and because the underlying
cause of the photometric variations being modeled remains unknown.
In practice, we also fit for the baseline flux
$f_0$ and an additive scalar jitter $s_{\text{TESS}}$. We fit the GP
hyperparameters using the parameter combinations
$\{ \ln{\omega_0}, \ln{S_0 \omega_0^4}, f_0, \log{s_{\text{TESS}}^2} \}$ with
uninformative priors.

The transit model within \texttt{exoplanet} fits the stellar mass
$M_s$, stellar radius $R_s$, and quadratic limb darkening parameters
$\{ u_1,u_2 \}$
along with the following planetary parameters: logarithmic orbital periods
$\ln{P}$, times of mid-transit $T_{0}$, logarithmic planet radii
$\ln{r_{p}}$, impact parameters $b$, and the eccentricity and argument of
periastron of LTT 3780c only; $\{e_c,\omega_c \}$.
We assume a circular orbit for the
inner planet LTT 3780b because its ultra-short period of 0.77 days implies
a very short circularization timescale of $\ll 1$ Myr \citep{goldreich66}.
Jointly fitting for
the physical stellar and planetary parameters enables us to derive the transit
observables $a/R_s$, $r_p/R_s$, and inclination $i$.
The joint GP plus two-planet transit model therefore includes 18 model
parameters:
$\{ f_0, \ln{\omega_0}, \ln{S_0 \omega_0^4}, \ln{s_{\text{TESS}}^2}, M_s, R_s,$
$u_1, u_2, \ln{P_b}, T_{0,b}, \ln{r_{p,b}}, b_b, \ln{P_c}, T_{0,c}, \ln{r_{p,c}}, b_c, e_c, \omega_c \}$.
\autoref{tab:priors} summarizes the \tess{} transit model parameter priors used
in this, our fiducial analysis.

\input{priors_exo}

Our full model is fit to the \tess{} PDCSAP light curve using the
\texttt{PyMC3} Markov Chain Monte-Carlo (MCMC) package  \citep{salvatier16}
implemented within \texttt{exoplanet}. We ran four simultaneous chains with 4000
tuning steps and 3000 draws in the final sample. \texttt{PyMC3} produces the
18-dimensional joint posterior probability density function (PDF) of the
model parameters. Median point estimates from the marginalized
posterior PDFs of the GP hyperparameters are used to construct the GP
predictive distribution whose mean function is shown in \autoref{fig:tess}
and is used to detrend the \tess{} light curve for visualization purposes.
Similarly, the median point estimates
of the transit model parameters are used to compute the `best-fit' transit
models shown in \autoref{fig:tess}.
\autoref{tab:tess} reports the median values of all model parameters
from their marginalized posterior PDFs along with their uncertainties from
the $16^{\text{th}}$ and $84^{\text{th}}$ percentiles.

\subsection{Precise radial-velocity analysis} \label{sect:rvs}
In our fiducial analysis,
we elected to fit the RVs independently of the transit data
but exploiting the strong priors on the orbital periods and mid-transit times
established by our \tess{} light curve
analysis (Sect.~\ref{sect:tessmcmc}). We note that
the information content within the \tess{} light curve and the RV measurements
with regards to their shared model parameters (i.e.
$\{ P_b, T_{0,b}, P_c, T_{0,c}, e_c, \omega_c \}$) is dominated by one
dataset or the other. In other words, the strongest constraints on each planet's
orbital period and mid-transit time are derived from the \tess{} and
ground-based
transit light curves. Conversely, most of the information regarding the
eccentricity and argument of periastron of LTT 3780c is derived from the RVs
since the planet's secondary eclipse is unresolved in the \tess{} light curve
and the eccentricity's effect on the transit duration is degenerate with
$a/R_s$, $r_p/R_s$, and $b$. Note that this is only an approximation as global
transit plus RV modeling can help to mitigate the eccentricity degeneracy
\citep{eastman19}. We will also consider a global model in
Sect.~\ref{sect:exofast}.

Although LTT 3780 is known to be relatively inactive, we do not expect its
surface to be completely static and homogeneous. As such, we expect some
temporally-correlated residual RV signals from magnetic activity that we model
with a quasi-periodic GP regression model for each spectrograph. The
quasi-periodic covariance kernel is

\begin{equation}
  k_{ij} = a^2 \exp{\left[ -\frac{(t_i-t_j)^2}{2\lambda^2} - \Gamma^2 \sin^2{\left( \frac{\pi |t_i-t_j|}{P_{\text{GP}}} \right)} \right]}
\end{equation}

\noindent and features four hyperparameters: the covariance amplitude $a$,
the exponential timescale $\lambda$, the coherence  $\Gamma$, and the periodic
timescale $P_{\text{GP}}$. We also fit
an additive scalar jitter $s_{\text{RV}}$ for each spectrograph to absorb any
excess white noise. Due to the
%chromatic dependence of magnetic activity signals and the
unique systematic noise
properties of each spectrograph, we fit a unique covariance amplitude and
scalar jitter to the data from each of the HARPS and HARPS-N
spectrographs. Throughout, the covariance parameters
$\{ \lambda, \Gamma, P_{\text{GP}} \}$, which only depend on signals originating
from the star, are kept fixed between the two spectrographs.

Our full RV model consists of a GP activity model for each spectrograph plus
independent Keplerian orbital solutions for each planet with RV
semi-amplitudes $K_b$ and $K_c$. We also fit for each spectrograph's systemic
velocity $\gamma$ to account for any RV offset between the two instruments.
Our full RV model therefore features 17 model 
parameters: $\{ \ln{a}_{\text{HARPS}}, \ln{a}_{\text{HARPS-N}}, \ln{\lambda}, \ln{\Gamma}, \ln{P_{\text{GP}}},$ $\ln{s}_{\text{HARPS}},\ln{s}_{\text{HARPS-N}}, \gamma_{\text{HARPS}}, \gamma_{\text{HARPS-N}},$ $P_b, T_{0,b}, \ln{K_b},$
$P_c, T_{0,c}, \ln{K_c}, h_c, k_c \}$ where $h_c=\sqrt{e_c}\cos{\omega_c}$ and
$k_c=\sqrt{e_c}\sin{\omega_c}$.
Note that the GP hyperparameters, scalar jitter parameters, and planetary
semi-amplitudes are fit in logarithmic units.
\autoref{tab:priors} includes each of the RV model parameter priors.

\begin{figure*}
  \centering
  \includegraphics[width=0.99\hsize]{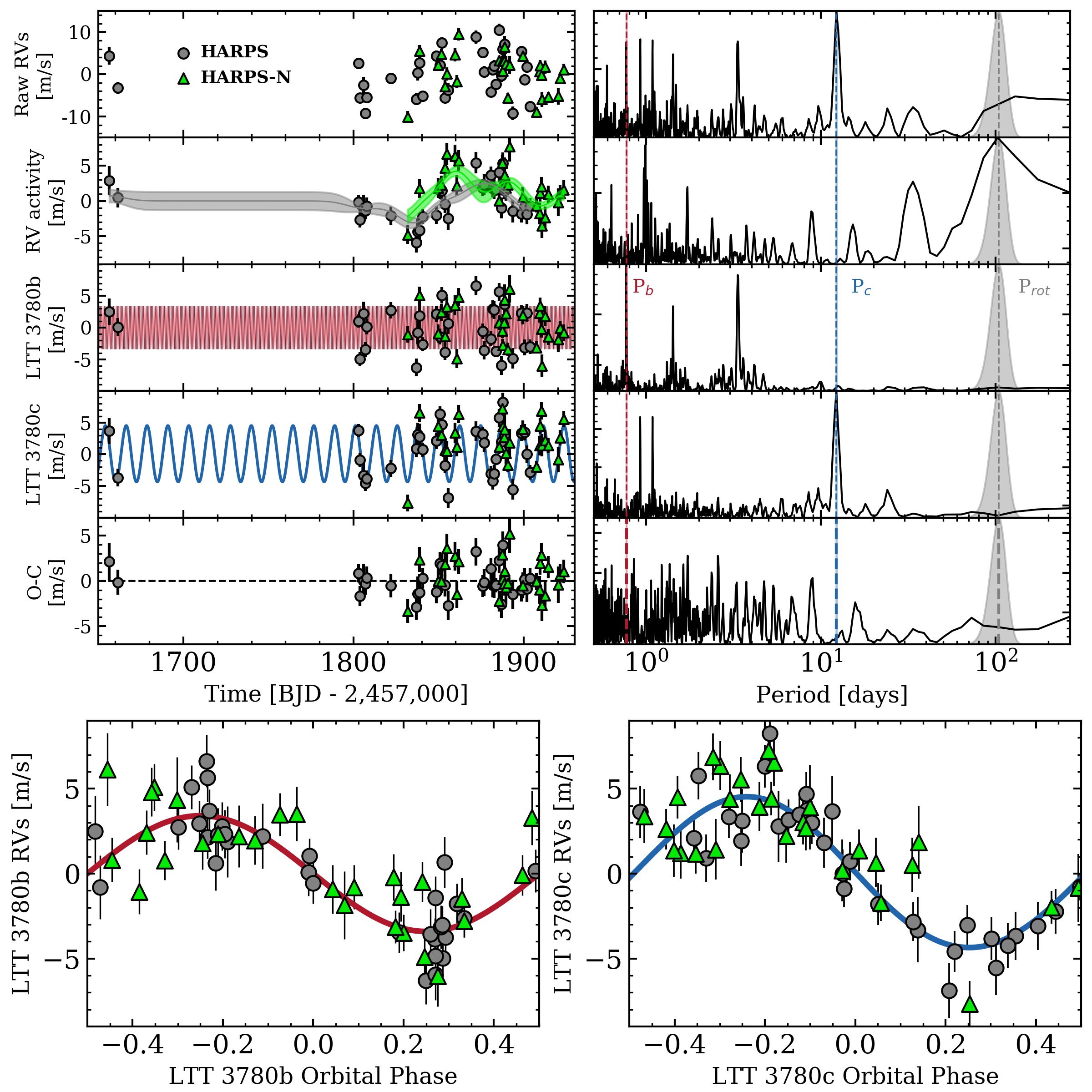}
  \caption{The RV data and individual model components from our analysis of the 
    HARPS (\emph{gray circles}) and HARPS-N (\emph{green triangles}) RV
    measurements. The data
    and models are depicted in the left column of the first five rows while
    their corresponding Bayesian generalized Lomb-Scargle periodograms are
    depicted in the right column. The marginalized posteriors of the LTT 3780b
    and c orbital periods are depicted as vertical lines along with the
    estimated stellar rotation period using the M dwarf activity-rotation
    relation from \cite{astudillodefru17b} (\prot{} $=104\pm 15$ days).
    \emph{First row}: the raw RV measurements.
    \emph{Second row}: the RV activity signal modeled with a quasi-periodic
    GP for each spectrograph.
    \emph{Third row}: the RV signal from LTT 3780b at 0.77 days.
    \emph{Fourth row}: the RV signal from LTT 3780c at 12.25 days.
    \emph{Fifth row}: the RV residuals.
    \emph{Bottom row}: the phase-folded RV signals of LTT 3780b and c.}
  \label{fig:rvs}
\end{figure*}

\autoref{fig:rvs} shows the raw RVs and the individual model components
including the RV activity along with LTT 3780b and c. The Bayesian generalized
Lomb-Scargle periodogram \citep[\texttt{BGLS};][]{mortier15} of each RV
component is also included in \autoref{fig:rvs}. 
The BGLS of the raw RVs exhibits a small number of significant peaks
(e.g. 3.1 days) that are
not strictly at either planet's orbital period. We will see that
the subtraction of the individual Keplerian orbits effectively removes these
periodicities such that they can
be attributed to harmonics of the planetary orbital periods.
The median RV model parameters from their marginalized posterior PDFs are used
to produce the models shown in \autoref{fig:rvs} and are reported in
\autoref{tab:tess} along with their $16^{\text{th}}$ and $84^{\text{th}}$
percentiles. The RV semi-amplitudes of LTT 3780b and c are found to be
$3.41^{+0.63}_{-0.63}$ and $4.44^{+0.82}_{-0.68}$ \mps{} and thus are
clearly detected at $5.4\sigma$ and $5.9\sigma$ respectively.
The resulting Keplerian RV signals are clearly
discernible in their phase-folded RV time series. The rms of the RV residuals
are found to be 1.55 and 1.74 \mps{} for HARPS and HARPS-N respectively.

M dwarfs are known to commonly host 2-3 planets per star out to 200 days
\citep[e.g.][]{dressing15a,ballard16,cloutier20,hardegree19} such that the
probability that a third planet exists around LTT 3780 is non-negligible.
However, the BGLS of the RV residuals in \autoref{fig:rvs} does not exhibit any
strong periodic signals that are statistically significant. This indicates that
a hypothetical third planet is unlikely to have been detected. To confirm this
robustly, we
considered a three-planet RV model, with fixed Keplerian parameters for LTT
3780b and c, plus a third Keplerian component `d' on a circular orbit. We
separately tested two three-planet models with differing priors on the orbital
period $P_d$: $\mathcal{U}(1.3,2.1)$ and
$\mathcal{U}(50,150)$ days. The chosen period limits approximately span
the two highest peaks in the BGLS of the RV residuals.
We then ran two separate MCMCs to sample the posteriors of the hypothetical
planet's
period, time of inferior conjunction (analogous to the mid-transit time), and
semi-amplitude. We find that neither model settles onto a preferred period or
phase and each marginalized $P_d$ posterior simply recovers its uninformative
prior. The lack of a well-defined maximum a-posteriori $P_d$ and $T_{0,d}$
prevents us from searching
the \tess{} light curve for any missed transit signals from the hypothetical
planet `d' and from placing a meaningful upper limit on the planet's mass.
We note that the only threshold crossing events identified
by the TPS in the \tess{} light curve were those corresponding to the confirmed
planets LTT 3780b and c.
Additionally, the recovered semi-amplitudes $K_d$ in both MCMCs favored zero
\mps{} with an upper limit of $K_d \leq 2.4$ \mps{} at 95\% confidence.
Taken together, these findings emphasize that the fiducial two-planet model for
the current dataset is likely complete as no third planet is detected
in our data.

\subsection{An alternative global transit + RV analysis} \label{sect:exofast}
To evaluate the robustness of the results derived in our fiducial analysis
(Sects.~\ref{sect:tessmcmc} and~\ref{sect:rvs}),
we conducted an independent analysis
using the \texttt{EXOFASTv2} exoplanet transit plus RV fitting package
\citep{eastman19}.
The methods of the \texttt{EXOFASTv2} fitting routine are detailed in
\cite{eastman19} although we provide a brief summary here.

To constrain the stellar-dependent parameters during the transit fit, we
feed \texttt{EXOFASTv2} the $M_s$ and $R_s$ parameter priors as in our fiducial
model. The routine also takes as input the pre-detrended light curves from
\tess{} and from ground-based facilities, and
performs a differential MCMC to evaluate the two-planet transit model
whose parameter priors are included in \autoref{tab:priors}.

There are a few notable differences between our fiducial analysis
(Sects.~\ref{sect:tessmcmc} and~\ref{sect:rvs}) and the \texttt{EXOFASTv2}
approach. The \texttt{exoplanet} model
simultaneously fits the hyperparameters of the GP detrending model plus the
transiting planet parameters to achieve
self-consistent detrending and transit models wherein the uncertainties in the
recovered planet parameters
are marginalized over uncertainties in the detrending model. Conversely,
\texttt{EXOFASTv2} uses pre-detrended light curves and so the aforementioned
marginalization of the planet parameter uncertainties over the GP
hyperparameters does not occur. Furthermore, the RV model in our fiducial
analysis includes the treatment of residual RV signals as a quasi-periodic GP
whereas, \texttt{EXOFASTv2} assumes the RV residuals to be well-represented by
a Gaussian noise term characterized by an additive jitter factor.

Our \texttt{EXOFASTv2} modeling 
has the important advantage of evaluating a global model that includes the
\tess{} light curve, ground-based transit light curves, and RV measurements.
The resulting planet parameters, including the orbital periods, mid-transit
times, eccentricities, and argument of periastron, will therefore be
self-consistent between all input datasets. In particular, each planet's
ephemeris will be more precisely constrained by the inclusion of the
ground-based transit light curves and the eccentricity of LTT 3780c will
be jointly constrained by its transit duration, Keplerian RV model, and the
stellar density. 
\texttt{EXOFASTv2} also fits a free dilution parameter to model any
discrepancies between the dilution applied to the PDCSAP light curve and the
true dilution.
%The inclusion of the additional datasets and the nature of the global modeling
%within \texttt{EXOFASTv2} results in an increased computational expense over
%our fiducial analysis strategy.

The results from our fiducial model in \autoref{tab:tess} are accompanied
by the results from our alternative analysis using \texttt{EXOFASTv2}.
We find consistency between the two models at $<1\sigma$ for nearly all model
parameters. This speaks to the robustness of the planetary model parameters
inferred from our data. The only exceptions are the $2\sigma$ and $2.8\sigma$
discrepant RV jitter parameters $s_{\text{HARPS}}$ and $s_{\text{HARPS-N}}$.
However, this is not alarming as the RV residuals, following the removal of the
two Keplerian solutions, are modeled with a GP in our fiducial model whereas the
\texttt{EXOFASTv2} model treats the residuals with a scalar jitter. Crucially,
these approaches yield consistent RV semi-amplitudes for
LTT 3780b and c whose agreement between the two models is $0.2\sigma$ and
$0.7\sigma$ respectively.

\section{Discussion} \label{sect:discussion}
\subsection{Fundamental planet parameters}
From our analysis of the \tess{} transit light curve we measure the planetary
radii of LTT 3780b and c to be $r_{p,b}=1.332^{+0.072}_{-0.075}$ \Rearth{} and
$r_{p,c}=2.30^{+0.16}_{-0.15}$ \Rearth{}. By
combining the \tess{} analysis with the
mid-transit times measured from transit follow-up observations,
we measure orbital periods for LTT 3780b and c to be
$P_b=0.7683881^{+0.0000084}_{-0.0000083}$ days and
$P_c=12.252048^{+0.000060}_{-0.000059}$ days. This places LTT 3780b at 0.012 AU
where it receives 106 times Earth's insolation. Assuming uniform heat
redistribution and a Bond albedo of zero, LTT 3780b has an equilibrium
temperature of $T_{\text{eq},b}=892$ K. Similarly, the orbital period of LTT
3780c places it at 0.077 AU where it receives 2.6 times Earth's insolation with
a zero-albedo equilibrium temperature of 353 K.

\begin{figure}
  \centering
  \includegraphics[width=0.99\hsize]{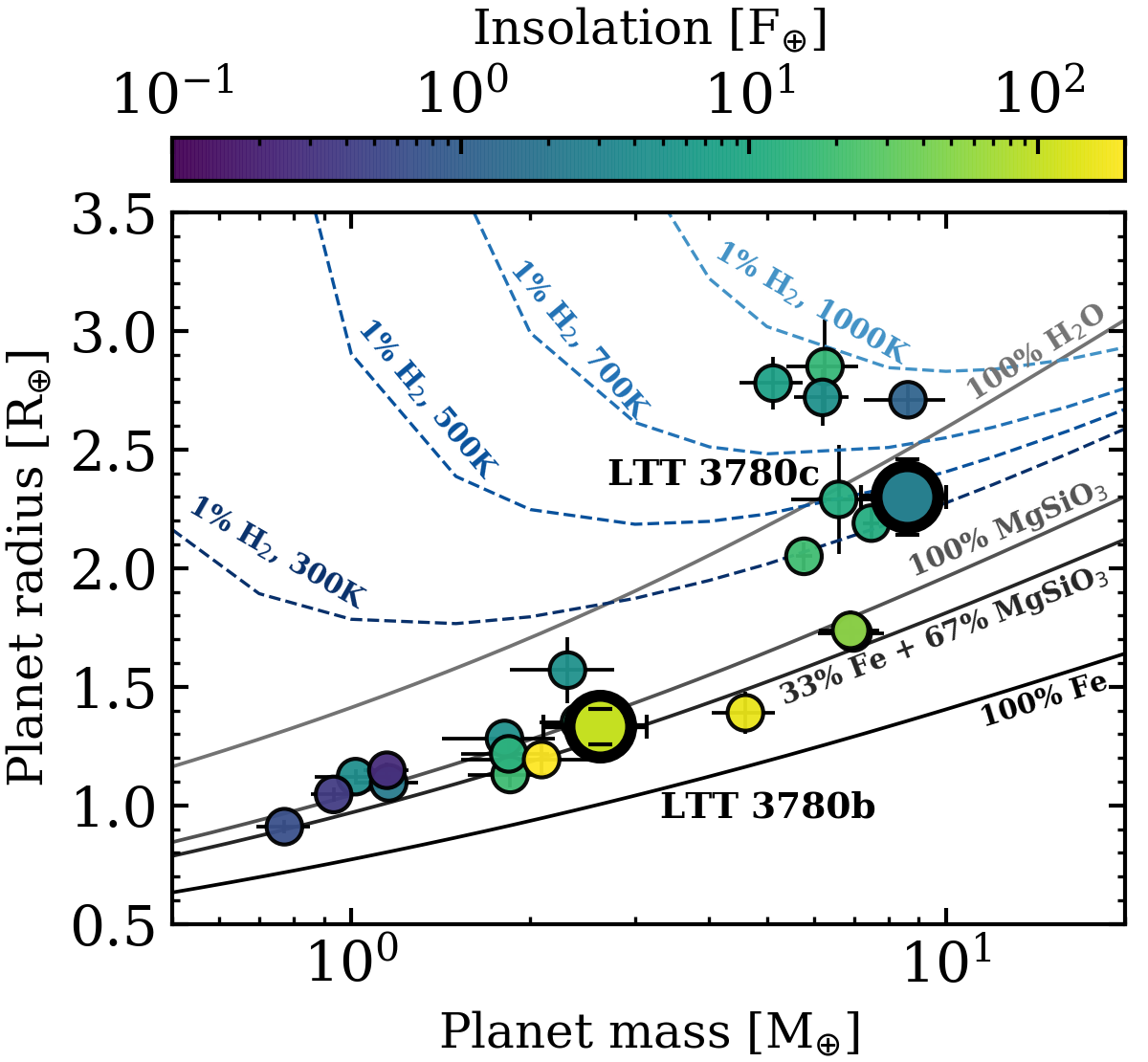}
  \caption{Planetary mass-radius diagram for small planets orbiting M dwarfs
    including LTT 3780b and c (\emph{bold symbols}).
    The \emph{solid curves} represent planetary internal structure models for
    bodies composed of 100\% water, 100\% silicate rock, 67\% rock plus 33\%
    iron (i.e. Earth-like), and 100\% iron by mass. The
    \emph{dashed curves} represent models of planets with Earth-like solid
    cores plus a 1\% by mass gaseous
    H$_2$ envelope at 1 mbar surface pressure and with the equilibrium
    temperature annotated next to each curve.
    Marker colors indicate the planet's insolation.}
  \label{fig:mr}
\end{figure}

From our RV analysis we measure planet masses of $m_{p,b}=2.62^{+0.48}_{-0.46}$
\Mearth{}
and $m_{p,c}=8.6^{+1.6}_{-1.3}$ \Mearth{,} which represent $5.6\sigma$ and
$5.9\sigma$ mass detections respectively.
By combining the planetary mass and radius measurements, we derive bulk
densities of $\rho_{p,b}=6.1^{+1.8}_{-1.5}$ g cm$^{-3}$ and
$\rho_{p,c}=3.9^{+1.0}_{-0.9}$ g cm$^{-3}$. \autoref{fig:mr} details the
mass-radius diagram of exoplanets around M dwarfs with masses
measured at the level of $\geq 3\sigma$, including the LTT 3780 planets.
The LTT 3780 planet masses and radii are compared to theoretical models of
fully-differentiated planetary interiors consisting of combinations of water,
silicate rock, and iron \citep{zeng13}.
In \autoref{fig:mr} we see that LTT 3780b is consistent with an Earth-like
bulk composition of 33\% iron plus 67\% magnesium silicate by mass. This
composition is shared by the majority of planets in the $\lesssim 1.5$
\Rearth{} size regime. We also consider models of
Earth-like solid cores that include 1\% H$_2$ envelopes by mass, over a range
of equilibrium temperatures from 300-1000 K \citep{zeng19}.
The mass and radius of LTT 3780c
appear consistent with a water-dominated bulk composition but also with a
predominantly Earth-like body that hosts an extended low mean molecular weight
atmosphere. Distinguishing between these two degenerate structure
models will require the extent of LTT 3780c's atmosphere to be investigated
through transmission spectroscopy. Due to the dependence of the atmospheric
scale height on the planet's surface gravity, the accurate interpretation of
forthcoming transmission spectroscopy observations will be facilitated by the
planetary mass measurements presented in this study. 
The feasibility of targeting LTT 3780c with transmission spectroscopy is
discussed in Sect.~\ref{sect:esmtsm}.

The LTT 3780 two-planet system adds to the growing number of confirmed
multi-planet systems around nearby M dwarfs with at least one transiting planet
(e.g. GJ 1132; \citealt{berta15,bonfils18},
K2-3; \citealt{crossfield15,damasso18},
K2-18; \citealt{montet15,cloutier19a},
L 98-59; \citealt{kostov19,cloutier19c},
LHS 1140; \citealt{dittmann17a,ment19},
LP 791-18; \citealt{crossfield19},
TOI-270; \citealt{gunther19},
TOI-700; \citealt{gilbert20,rodriguez20},
TRAPPIST-1; \citealt{gillon17a}).
With their sub-Neptune-sized radii and measured masses presented herein, both 
LTT 3780b and c contribute directly to the completion of the \tess{} 
level one science requirement to obtain masses for fifty planets
smaller than 4 \Rearth{.}

\subsection{Implications for the origin of the radius valley around mid-M
  dwarfs}
The occurrence rate distribution of close-in planet radii around Sun-like stars
features a bimodality with a dearth of planets at $1.7-2.0$ \Rearth{}
known as the radius valley \citep{fulton17,mayo18}. This feature likely results
from the existence of a transition between predominantly rocky planets and
larger planets that host significant H/He envelopes, as a function of planet
radius and orbital separation. The slope of the radius valley in $P-r_p$ space
marks the critical radius separating rocky and non-rocky planets as a function
of orbital period. The empirical slope of the radius valley
around Sun-like stars is consistent with models of thermally-driven
atmospheric mass loss such as photoevaporation and core-powered mass loss
\citep{vaneylen18,martinez19,wu19}.
However for mid-K to mid-M dwarfs, the radius valley slope flattens and
becomes increasingly favored by models of an alternative formation pathway for
terrestrial planets in a gas-poor environment \citepalias{cloutier20}.
%Investigations of the close-in planet population with \emph{K2} have also
%revealed tentative variations in the radius valley structure with

\begin{figure}
  \centering
  \includegraphics[width=0.99\hsize]{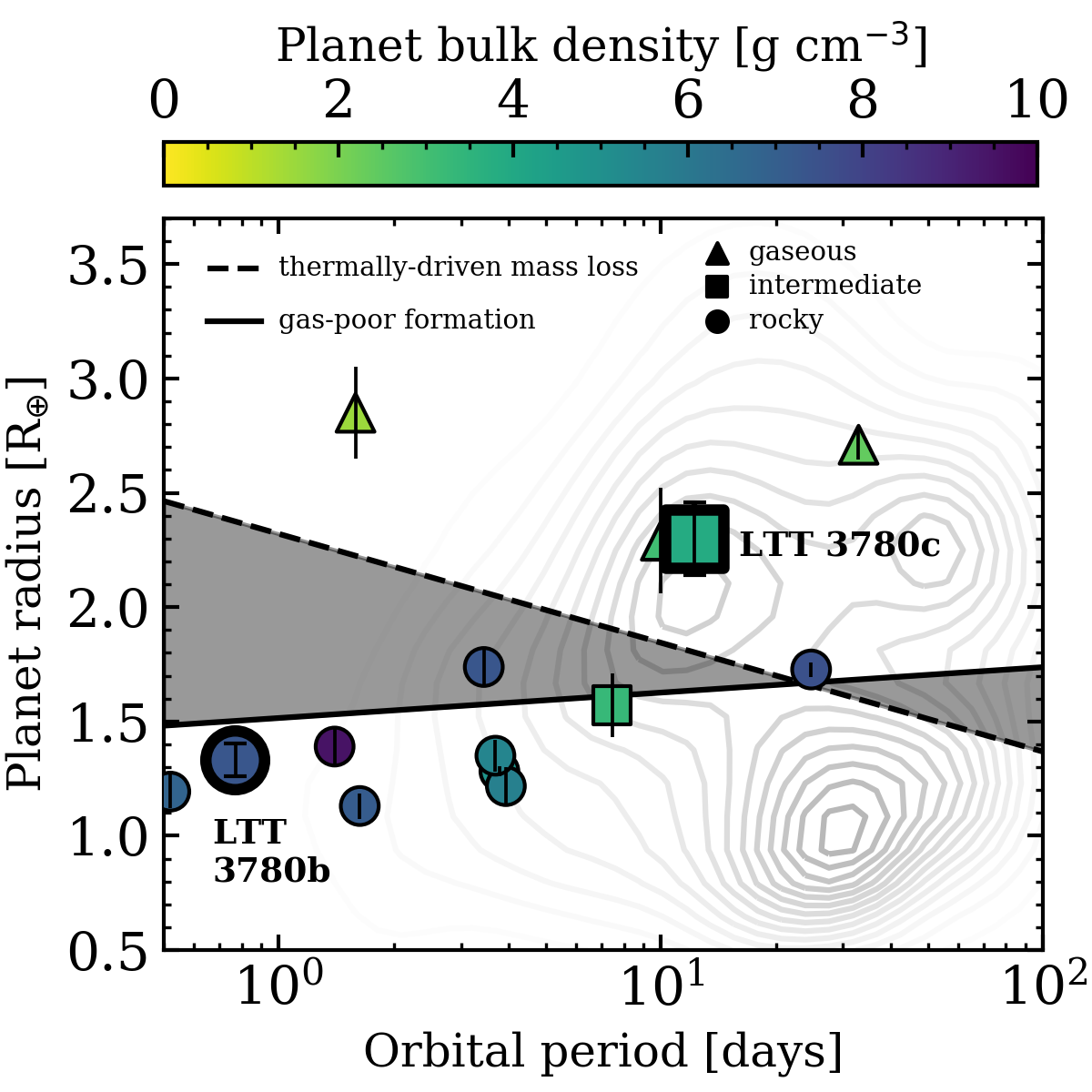}
  \caption{Period, radii, and bulk densities of M dwarf planets with
    precise RV masses compared to the empirical location of the radius valley
    around low mass stars versus orbital period and planet radius. LTT 3780b
    and c are depicted with the \emph{bold symbols}. Contours represent the
    planetary occurrence rates around low mass stars \citepalias{cloutier20}.
    Planet marker shapes depict the planet's compositional disposition as either
    rocky (\emph{circles}), gaseous (\emph{triangles}), or intermediate
    (\emph{squares}). Marker colors indicate the planet's bulk density.
    The \emph{dashed} and \emph{solid} lines depict the locations of the radius
    valley around low mass stars from model predictions of thermally-driven
    atmospheric mass loss and from gas-poor terrestrial planet formation
    respectively. The \emph{shaded regions} highlight where the
    model predictions of planetary bulk compositions are discrepant between the
    two models.}
  \label{fig:radval}
\end{figure}

\autoref{fig:radval} depicts the LTT 3780 planets in $P-r_p$ space, along
with the subset of M dwarf planets from \autoref{fig:mr} with RV-derived masses.
The planets in \autoref{fig:radval} are classified as having a bulk
composition that is either rocky, gaseous, or intermediate based on their mass
and radius. Rocky planets are defined as planets that are consistent with
having a bulk density greater than that of 100\% MgSiO$_3$ given their size.
Similarly, unambiguously gaseous planets are defined as planets that are
consistent with having a bulk density less than that of 100\% H$_2$O given their
size. The remaining planets are flagged as having bulk compositions that are
intermediate between rocky and gaseous.
LTT 3780b and c have rocky and intermediate dispositions respectively
(\autoref{fig:mr}).

In \autoref{fig:radval}, LTT 3780b and c are shown to
span the empirically-derived location of the radius valley around low mass
stars under the gas-poor formation and photoevaporation models
\citepalias{cloutier20}. The slope of the radius valley around low mass stars is
considerably flatter than around Sun-like stars, with the former slope being
consistent with gas-poor formation while the latter is
more consistent with a thermally-driven atmospheric mass loss process. 
To compare the compositions of planets around low mass stars to the
rocky/non-rocky transition locations in \autoref{fig:radval}, we scale the
transition measured around Sun-like stars down to the low stellar mass regime
under the photoevaporation model
\citep[$r_p \propto (M_s/M_{\odot})^{1/4}$;][]{wu19}\footnote{The median stellar
  mass in the sample of Sun-like stars from \cite{martinez19} is 1.01 \Msun{.}
  The median stellar mass in the sample of low mass stars from
  \citetalias{cloutier20}
  is 0.65 \Msun{.} The resulting scaling of the rocky/non-rocky transition
  from Sun-like stars to the low stellar mass regime under photoevaporation is
  $(0.65/1.01)^{1/4}=0.896$ \citep{wu19}.}.
The slope measured around low mass stars is plotted verbatim in
\autoref{fig:radval}. Both models predict that LTT 3780b should have
a rocky bulk composition in which any residual gaseous envelope only
contributes marginally to the planet's mass and radius. Indeed these
predictions are consistent with our finding that LTT 3780b has an Earth-like
composition. Similarly, both models predict that LTT 3780c should be non-rocky
in that it
should have retained a substantial gaseous envelope and therefore be
inconsistent with having a bulk rocky composition. Although we cannot
definitively identify the bulk composition of LTT 3780c with our data,
due to internal structure model degeneracies,
we confirm that LTT 3780c is consistent with both model predictions. In other
words, the models correctly identify LTT 3780c as being inconsistent with an
Earth-like composition and requires a significant
amount of volatile material or H/He gas to explain its mass and radius.

\subsubsection{Planetary mass limits from photoevaporation models} \label{sect:pe}
Stars such as LTT 3780 with multi-transiting planets that span the
radius valley provide valuable test cases of radius valley emergence models.
The virtue of these systems is that limits on the planetary masses can be
derived by scaling the properties of one planet to the other \citep{owen20}.
For example, 
assuming that the initial H/He envelope of the rocky planet below the valley
has been completely stripped by some physical process, 
the theoretical minimum mass of the non-rocky planet above the valley can be
calculated by scaling its properties to those of the rocky planet. An
equivalent principle can be used to derive the maximum mass of the rocky
planet. The power
of this comparative scaling of planets within the same planetary system is
that certain unobservable quantities that directly affect final planet masses
are scaled out. An example of this is the host star's XUV luminosity history in
the photoevaporation scenario \citep{owen20}.

A full derivation is presented in \autoref{app:pe} but here we simply
state the condition for the consistency of the gaseous (i.e. non-rocky) and
rocky planet parameters with the photoevaporation model. This requires that
the gaseous planet's mass loss timescale exceeds the maximum mass loss
timescale of the rocky planet \citep{owen20}. This condition leads to

\begin{equation}
  1\leq \frac{m_{\text{core,gas}}^{0.64}}{m_{\text{core,rock}}}
    \left( \frac{a_{\text{gas}}}{a_{\text{rock}}} \right)^{2/3}
    r_{\text{core,rock}}^{4/3}. \label{eq:ineqpe}
\end{equation}

\noindent where each planet's core mass and radius are given in units of the
Earth. In the LTT 3780 system we define LTT 3780b to be the rocky planet below
the valley whose H/He envelope has been photoevaporated away leaving behind
a solid core whose mass and radius are equal to the planet's total mass and
radius: $m_{\text{core,rock}}=m_{p,b}=2.62\pm 0.47$ \Mearth{} and
$r_{\text{core,rock}}=r_{p,b}=1.332\pm 0.074$ \Rearth{.} The gaseous planet above
the valley is then LTT 3780c, whose mass is assumed to be dominated by
an Earth-like core such that $m_{\text{core,gas}}=m_{p,c}=8.6\pm 1.5$ \Mearth{} and
whose core radius is approximated by the mass-radius relation for Earth-like
bodies \citep[$r_p\propto m_p^{1/3.7}$;][]{zeng16}. Lastly, the semimajor axes
$a_{\text{rock}}$ and $a_{\text{gas}}$ are $a_b= 0.01211\pm 0.00012$ AU and
$a_c= 0.07673\pm 0.00076$ AU respectively.

Using \autoref{eq:ineqpe} and sampling the planetary parameters
$\Theta = \{ m_{p,b}, a_b, r_{p,b}, a_c \}$
from their marginalized posterior PDFs, we find that the mass of LTT 3780c
must be $\geq 0.49\pm 0.15$ \Mearth{}
in order to be consistent with the photoevaporation model.
In the same way, but by replacing $m_{p,b}$ with $m_{p,c}$ in the set
$\Theta$, we calculate that the mass of LTT 3780b must be
$\leq 19.6\pm 2.8$
\Mearth{} to be consistent with photoevaporation. Clearly the measured masses
$m_{p,c}=8.6\pm 1.5$ \Mearth{}
and $m_{p,b}=2.62\pm 0.47$ \Mearth{} are both consistent
with predictions from the photoevaporation model, implying that photoevaporation
is a feasible process for sculpting the observed architecture of the LTT 3780
system.

A few notable caveats exist with the planetary mass limits imposed by the
photoevaporation model in \autoref{eq:ineqpe} \citep{owen20}.
These are discussed in \autoref{app:pe}.

\subsubsection{Planetary mass limits from core-powered mass loss models}
Similarly to the photoevaporation model, we can compare the mass loss timescales
of the LTT 3780 planets under the core-powered mass loss scenario
\citep{ginzburg18,gupta19,gupta20} to constrain
their permissible planet masses under that model. In the core-powered mass loss
scenario, the lower atmosphere is in thermal contact with the planetary core
which conducts energy from its formation into the atmosphere. This
heat flux drives convective heat transport radially outwards to the
radiative-convective boundary (RCB) of the atmosphere, above which
the atmosphere is isothermal at \teq{} and atmospheric
cooling is radiative. The physical
limit to the atmospheric mass loss rate is given by the thermal velocity of the
gas at the Bondi radius; the radial distance at which the escape velocity equals
the thermal sound speed $c_s = \sqrt{k_B T_{\text{eq}}/\mu}$ where $k_B$ is the
Boltzmann constant and $\mu$ is the atmospheric mean molecular weight which we
fix to 2 amu for H$_2$.

The derivation of the mass loss timescale in the core-powered mass loss model
is presented in \autoref{app:cpml}. As in the photoevaporation scenario, we
require that the mass loss timescale for the gaseous planet exceeds that of the
rocky planet which leads to the following condition for consistency of the
planetary parameters with the core-powered mass loss model:

\begin{align}
  1 &\leq \left( \frac{m_{\text{core,gas}}}{m_{\text{core,rock}}} \right)
  \left( \frac{T_{\text{eq,gas}}}{T_{\text{eq,rock}}} \right)^{-3/2} \nonumber \\
  & \exp{\left[ c' \left(\frac{m_{\text{core,gas}}}{T_{\text{eq,gas}}\: r_{p,\text{gas}}} - \frac{m_{\text{core,rock}}}{T_{\text{eq,rock}}\: r_{p,\text{rock}}} \right) \right]},
  \label{eq:ineqcpml}
\end{align}

\noindent where the constant $c' =G\mu/k_B\approx 10^{4}$ \Rearth{} K
\Mearth{$^{-1}$},
$T_{\text{eq,gas}}=T_{\text{eq},c}=323\pm 16$ K, 
$T_{\text{eq,rock}}=T_{\text{eq},b}=816\pm 40$ K,
$r_{p,\text{gas}}=r_{p,c}=2.30\pm 0.16$ \Rearth{,} and  
$r_{p,\text{rock}}=r_{p,b}=1.332\pm 0.074$ \Rearth{.}
The inequality in \autoref{eq:ineqcpml} has no analytic solution so we solve
for the limiting masses of $m_{\text{core,gas}}$ and $m_{\text{core,rock}}$ by again
sampling the planetary parameters
$\{ m_{\text{core,rock}}, T_{\text{eq,rock}}, r_{p,\text{rock}}, m_{\text{core,gas}}, T_{\text{eq,gas}}, r_{p,\text{gas}} \}$
from their marginalized posterior PDFs and numerically
solving for the limiting core masses. Recall that both planets are assumed to
have small envelope mass fractions such that $m_{\text{core}}\approx m_p$. 

Under the core-powered mass loss mechanism, we find that the mass of LTT 3780c
must be $\geq 2.1\pm 0.5$ \Mearth{} to be consistent with the model.
Similarly, by
solving for $m_{\text{core,rock}}$ we calculate that the mass of LTT 3780b must be
$\leq 12.6\pm 2.9$ \Mearth{.} As with the photoevaporation mass limits from
Sect.~\ref{sect:pe}, the measured masses
$m_{p,c}=8.6\pm 1.5$ \Mearth{} and $m_{p,b}=2.62\pm 0.47$ \Mearth{} are both
consistent with predictions from the core-powered mass loss model.

The masses of LTT 3780b and c recovered in this study from HARPS and HARPS-N
RV measurements are both consistent with radius valley emergence model
predictions from photoevaporation and core-powered mass loss, two
physical processes that thermally drive atmospheric escape on close-in
planets. Thus, the recovered masses of LTT 3780b and c are unable to provide
strong evidence for the inapplicability of either mechanism. However, the
photoevaporation and core-powered mass loss models do make distinct predictions
for the maximum mass of the rocky planet and the minimum mass of the non-rocky
in systems like LTT 3780 that feature such planet pairs.
Therefore, other systems with multi-transiting planets that span the
radius valley may exist for which either photoevaporation or core-powered
mass loss may be ruled out by the planets' masses. This prospect is especially
viable for increasingly compact systems wherein the ratios
$a_{\text{gas}}/a_{\text{rock}}$ and $T_{\text{eq,gas}}/T_{\text{eq,rock}}$ approach
unity.

\subsubsection{Planetary mass limits from gas-poor terrestrial planet
  formation models}
Unlike the photoevaporation and core-powered mass loss scenarios, it
is not clear that analogous arguments can be made within the gas-poor
formation framework to scale out unknown system parameters and
place limits on the permissible planet masses. This is because the model
invokes the formation of two
planet populations with distinct rocky and non-rocky bulk compositions
\citep{lee14,lee16,lopez18}. Both populations are thought to form cores of
rock and ice
but only the non-rocky population subsequently accretes a gaseous envelope
prior to disk dispersal after a few Myrs \citep{haisch01,cloutier14}.
Because the gas accretion term only impacts the non-rocky planet population,
unobservable quantities for the LTT 3780 system when it was just a few Myrs old,
such as the local density of the gaseous disk, the disk structure, and the disk
dynamics, cannot be scaled out by comparing the rocky and non-rocky planet
parameters. As such,
we are not in a position to compare the LTT 3780 planet masses to constraints
imposed by the gas-poor terrestrial planet formation model.

\subsection{TTV analysis}
We used the \texttt{TTV2Fast2Furious python} package \citep{hadden19}
to predict the amplitudes
of transit timing variations (TTVs) of the planets LTT 3780b and c. We ran
$10^3$ realizations with the planetary masses being sampled from their
marginalized posterior PDFs from our RV analysis (Sect.~\ref{sect:rvs}). The
stellar mass, planet orbital periods, and times of mid-transit are drawn from
their respective priors used in our RV analysis. Recall that the free
eccentricity of LTT 3780b is assumed to be zero because of its short
circularization timescale. Furthermore, due to their large period
ratio ($P_b=0.768388$ days, $P_c=12.252048$ days, $P_c/P_b=15.945130$),
imposing a non-zero free eccentricity on either planet will have a
negligible effect on their TTV amplitudes so we fix the input free
eccentricities to zero. The forced eccentricities induced by the planets'
mutual interactions are calculated within \texttt{TTV2Fast2Furious}. Arguments
of periastron are drawn from $\mathcal{U}(0,2\pi)$.

In each realization, with its unique set of parameters, we compute each planet's
maximum deviation from a linear ephemeris over a 2-year baseline beginning with
the commencement of the \tess{} sector 9 observations.
%and the most recent ground-based transit observation
%taken on UT February 9, 2020.
Over the $10^3$ realizations we find maximum TTV amplitudes of 0.02 and
1 second for LTT 3780b and c respectively. The small amplitude of the expected
TTV signals make the LTT 3780 system a poor candidate for intensive transit
follow-up to derive TTV masses of the two known planets. However, ongoing
transit observations of LTT 3780c may reveal TTVs induced by an insofar unseen
outer planet. For this purpose, we note that LTT 3780 is scheduled to be
observed in sector 35 of the \tess{} extended mission between UT February 9 and
March 7, 2021.

\subsection{Prospects for atmospheric characterization} \label{sect:esmtsm}
The stellar and planetary parameters of the LTT 3780 system make the planets
LTT 3780b and c accessible targets for atmospheric characterization via
emission and transmission spectroscopy respectively. Assuming
uniform heat redistribution and a Bond albedo of zero, the equilibrium
temperature of LTT 3780c is $T_{\text{eq},c}=353$ K. The expected depth of its
transmission features up to two atmospheric scale heights
\citep{stevenson16,fu17}, in a cloud-free low mean
molecular weight atmosphere ($\mu=2$), is 79 ppm. Alternatively, it is expected
that some mini-Neptune atmospheres are metal enriched \citep{fortney13}
which will partially suppress transmission feature depths to 32 ppm in a
100x solar metallicity atmosphere ($\mu \approx 5$).
Simulated transit observations with \texttt{PandExo} \citep{batalha17} confirm
that molecular features in a clear, low mean molecular weight atmosphere
will be detectable at $\geq 5\sigma$ confidence from a single transit
observation with \emph{JWST}/NIRISS slitless spectroscopy\footnote{Note that
  LTT 3780's $J$-band magnitude of 9.007 does not exceed any imposed brightness
  limit in the NIRISS Single Object Slitless Spectroscopy (SOSS) mode.}
\citep{kreidberg15a}. Four transits would be required to reach a similar
precision for a 100x solar metallicity atmosphere. We also note the caveat that
if high altitude clouds are present on LTT 3780c, as seen for many other planets
in its size regime \citep{crossfield17}, additional observing time will be
required.

\begin{figure*}
  \centering
  \includegraphics[width=0.98\hsize]{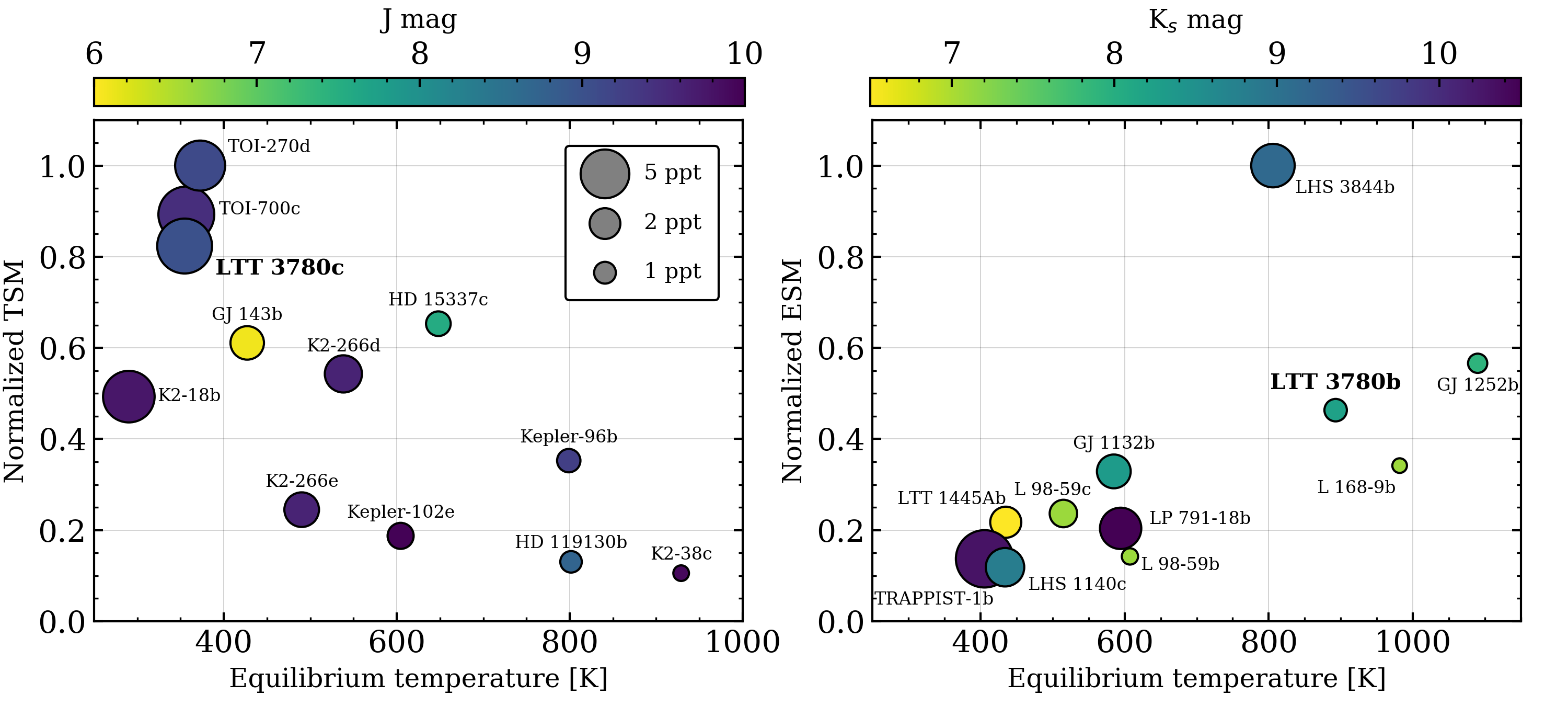}
  \caption{Normalized atmospheric characterization metrics \citep{kempton18}
    versus equilibrium temperature and host star apparent magnitude.
    \emph{Left panel}: the transmission spectroscopy metric (TSM) for warm
    mini-Neptunes around bright host stars ($J<10$) with
    $P \in [10,40]$ days and $r_p \in [2,3]$ \Rearth{,} including LTT 3780c.
    Marker colors depict the host star's $J$-band magnitude.
    \emph{Right panel}: the emission spectroscopy metric (ESM) for favorable
    close-in rocky planets ($r_p<1.5$ \Rearth{)} including LTT 3780b.
    Marker colors depict the host star's $K_s$-band magnitude.
    In both panels the marker sizes depict the primary transit depths.}
  \label{fig:esmtsm}
\end{figure*}

For LTT 3780c, we can also consider the
transmission spectroscopy metric \citep[TSM;][]{kempton18} which is
proportional to the expected S/N of transmission features in a cloud-free
atmosphere. Based on the TSM, LTT 3780c is among the best warm
mini-Neptunes ($P \in [10,40]$ days, $r_p \in [2,3]$ \Rearth{)} for atmospheric
characterization via transmission spectroscopy observations. To date, the best
such planets are the \tess{-}discovered planets TOI-700c
\citep{gilbert20,rodriguez20},
TOI-270d \citep{gunther19}, and LTT 3780c, whose TSM values are all within 17\%
of each other and are at minimum 17\% greater than that of the next best
potential target:
HD 15337c \citep{dumusque19}. The TSM values of favorable warm mini-Neptunes
are reported in \autoref{tab:tsm} and are compared in \autoref{fig:esmtsm}.

\input{tsmtable}

The ultra-short period planet LTT 3780b has a zero-albedo equilibrium
temperature of
$T_{\text{eq},b}=892$ K. The hot dayside of LTT 3780b makes it a very attractive
target for atmospheric characterization via emission
spectroscopy observations. In particular, eclipse observations can help to
discern whether the planet has
retained a substantial atmosphere or if its emitting temperature is
consistent with that of pure rock. The distinction between a 1 bar atmosphere
and a bare rocky surface on LTT 3780b will be accessible with a single
\emph{JWST}/MIRI eclipse observation \citep{koll19}.

Similarly to the TSM, the expected S/N of thermal emission
signatures at 7.5 $\mu$m is  
proportional to the emission spectroscopy metric \citep[ESM;][]{kempton18}.  
Computing the ESM for hot planets with likely terrestrial compositions
($r_p<1.5$ \Rearth{)}, that are
favorable targets for emission spectroscopy measurements, reveals that LTT 3780b
is among the best such planets (\autoref{tab:esm}, \autoref{fig:esmtsm}).
The ESM for LTT 3780b is the third highest among these planets and
closely matches that of GJ 1252b \citep{shporer20}. Both of these targets have
ESM values that are nearly half that of LHS 3844b \citep{vanderspek19}, a
rocky planet whose thermal phase curve has been characterized
by the \emph{Spitzer Space Telescope} and found to be consistent with a dark
basaltic surface that lacks any substantial atmosphere \citep{kreidberg19}.

\input{esmtable}

The favorable ESM and TSM values of LTT 3780b and c respectively make them both
accessible targets for atmospheric characterization. Together they present a
unique opportunity to conduct direct comparative studies of exoplanet
atmospheres among planets within the same planetary system which is critical
for informing our understanding of the formation and evolution of close-in
planets at a range of sizes and equilibrium temperatures.

\subsection{An independent analysis of the LTT 3780 system by CARMENES}
Following the announcement of the planet candidates TOI-732.01 and 02
in May 2019, multiple PRV instrument teams began working
towards the mass characterization of these potential planets. This study has
presented the subset of those efforts from HARPS and HARPS-N but
we acknowledge that the CARMENES team has also submitted a paper presenting
their own RV time series and analysis \citep{nowak20}.
Although the submissions of these
complementary studies were coordinated between the two groups, their respective
data, analyses, and writeups were intentionally conducted independently.
%This approach was
%designed with the goal of either confirming the accuracy of the measured planet
%parameters or to identify sources of inconsistencies stemming from 
%differences between the two datasets or differences in the data analysis
%procedures. Because we have not seen the results of the CARMENES study,
%%As such, we are unable to comment on their findings at this time.

\section{Summary} \label{sect:summary}
In this study, we present the LTT 3780 multi-transiting system from the \tess{}
mission. The newly discovered planets LTT 3780b and c are 
confirmed with intensive follow-up observations that includes ground-based
transit photometry, reconnaissance spectroscopy, high-resolution imaging, and
63 precise
RV measurements from HARPS and HARPS-N. Our main findings are summarized below.

\begin{itemize}
\item LTT 3780 is a bright ($V=13.07$, $K_s=8.204$) mid-M dwarf with
  $M_s=0.401\pm 0.012$ \Msun{} and $R_s=0.374\pm 0.011$ \Rsun{,} located at
  22 pc.
\item LTT 3780b is a hot rocky exoplanet with $P_b=0.77$ days,
  $r_{p,b}=1.33\pm 0.07$ \Rearth{,} and $m_{p,b}=2.62^{+0.48}_{-0.46}$ \Mearth{,} making
  its bulk composition consistent with that of the Earth.
\item LTT 3780c is a warm mini-Neptune with $P_c=12.25$ days,
  $r_{p,c}=2.30\pm 0.16$ \Rearth{,} and $m_{p,c}=8.6^{+1.6}_{-1.3}$ \Mearth{.}
  Its bulk composition is inconsistent with being Earth-like and requires a
  significant amount of volatile material or H/He gas to explain its mass and
  radius.
\item The two planets span the radius valley around low mass stars which enables
  the comparison of their planetary parameters to predictions from models of the
  emergence of the radius valley. Both planets' physical and orbital properties
  are shown to be
  consistent with predictions of atmospheric escape from photoevaporation and
  from core-powered mass loss.
\item The brightness and small size of LTT 3780 make the planets LTT 3780b and c
  accessible targets for atmospheric characterization of a hot rocky planet and
  a warm mini-Neptune via emission and transmission spectroscopy observations
  respectively.
\end{itemize}

\acknowledgements
RC is supported by a grant from the National Aeronautics and Space
Administration in support of the TESS science mission.
We thank Amber Medina for assistance with detrending the \tess{} light curve.
We thank Sam Hadden for discussions regarding the TTV analysis.
We also thank the anonymous referee for their comments that helped to improve
the completeness of our paper.

NAD acknowledges the support from FONDECYT 3180063.

AM acknowledges support from the senior Kavli Institute Fellowships.

JGW is supported by a grant from the John Templeton Foundation. The opinions
expressed in this publication are those of the authors and do not necessarily
reflect the views of the John Templeton Foundation.

CZ is supported by a Dunlap Fellowship at the Dunlap Institute for Astronomy
\& Astrophysics, funded through an endowment established by the Dunlap family
and the University of Toronto.

IJMC acknowledges support from the NSF through grant AST-1824644,
and from NASA through Caltech/JPL grant RSA-1610091. 

FL gratefully acknowledges a scholarship from the Fondation Zd\u{e}nek
et Michaela Bakala.

MS thanks the Swiss National Science Foundation (SNSF) and the Geneva
University for their continuous support to our exoplanet researches. This work
has been in particular carried out in the frame of the National Center for
Competence in Research `PlanetS' supported by SNSF.

CAW acknowledges support from Science and Technology Facilities Council grant
ST/P000312/1.

NCS acknowledges supported by FCT - Funda\c{c}\~ao para a Ci\^encia e a
Tecnologia through national funds and by FEDER through COMPETE2020 - Programa
Operacional Competitividade e Internacionalização by these grants:
UID/FIS/04434/2019; UIDB/04434/2020; UIDP/04434/2020; PTDC/FIS-AST/32113/2017
\& POCI-01-0145-FEDER-032113; PTDC/FIS-AST/28953/2017 \&
POCI-01-0145-FEDER-028953.

MPi gratefully acknowledges the support from the European Union Seventh
Framework Programme (FP7/2007-2013) under Grant Agreement No. 313014 (ETAEARTH).

JRM acknowledges the support by the CAPES, CNPq, and FAPERN Brazilian agencies.

This work has been partially supported by the National Aeronautics and Space
Administration under grant No.  NNX17AB59G issued through the Exoplanets
Research Program.

We acknowledge the use of public TESS Alert data from the pipelines at the
TESS Science Office and at the TESS Science Processing Operations Center.

This work makes use of observations acquired with the T150 telescope at 
Sierra Nevada Observatory, operated by the Instituto de Astrof\'isica de 
Andalucía (IAA-CSIC).

The MEarth Team gratefully acknowledges funding from the David and Lucile
Packard Fellowship for Science and Engineering (awarded to D.C.). This material
is based upon work supported by the National Science Foundation under grants
AST-0807690, AST-1109468, AST-1004488 (Alan T. Waterman Award), and AST-1616624.
This work is made possible by a grant from the John Templeton Foundation. The
opinions expressed in this publication are those of the authors and do not
necessarily reflect the views of the John Templeton Foundation. This material
is based upon work supported by the National Aeronautics and Space
Administration under Grant No. 80NSSC18K0476 issued through the XRP Program.

This publication makes use of data products from the Two Micron All Sky Survey,
which is a joint project of the University of Massachusetts and the Infrared
Processing and Analysis Center/California Institute of Technology, funded by
the National Aeronautics and Space Administration and the National Science
Foundation.

This work has made use of data from the European Space Agency (ESA) mission
\gaia{} (\url{https://www.cosmos.esa.int/gaia}), processed by the \gaia{} Data
Processing and Analysis Consortium (DPAC,
\url{https://www.cosmos.esa.int/web/gaia/dpac/consortium}). Funding for the
DPAC has been provided by national institutions, in particular the institutions
participating in the \gaia{} Multilateral Agreement.

This work makes use of observations from the LCOGT network.

Resources supporting this work were provided by the NASA High-End Computing
(HEC) Program through the NASA Advanced Supercomputing (NAS) Division at Ames
Research Center for the production of the SPOC data products.

This work was supported by the French National Research Agency
in the framework of the Investissements d’Avenir program
(ANR-15-IDEX-02), through the funding of the "Origin of Life" project of
the Univ. Grenoble-Alpes.

\facilities{TESS, TRES, LCOGT, OSN, TRAPPIST-North, MEarth-North, SOAR,
  Gemini/NIRI, ESO 3.6m/HARPS, TNG/HARPS-N.}

\software{\texttt{AstroImageJ} \citep{collins17},
  \texttt{astropy} \citep{astropyi,astropyii},
  \texttt{BANZAI} \citep{mccully18},
  \texttt{batman} \citep{kreidberg15},
  \texttt{BGLS} \citep{mortier15},
  \texttt{celerite} \citep{foremanmackey17},
  \texttt{emcee} \citep{foremanmackey13},
  \texttt{EvapMass} \citep{owen20},
  \texttt{EXOFAST} \citep{eastman13},
  \texttt{EXOFASTv2} \citep{eastman19},
  \texttt{exoplanet} \citep{foremanmackey19},
  \texttt{forecaster} \citep{chen17},
  \texttt{PandExo} \citep{batalha17},
  \texttt{PyMC3} \citep{salvatier16},
  \texttt{scipy} \citep{scipy},
  \texttt{STARRY} \citep{luger19},
  \texttt{Tapir} \citep{jensen13},
  \texttt{TERRA} \citep{anglada12},
  \texttt{TTV2Fast2Furious} \citep{hadden19}.}

\bibliographystyle{apj}
\bibliography{refs}
\input{tesstable_exofast}
\input{appendices}

\suppressAffiliationsfalse
\allauthors
\end{document}

%% file: allauthors.tex
\author[0000-0001-5383-9393]{Ryan Cloutier}
\affiliation{Center for Astrophysics $|$ Harvard \& Smithsonian, 60 Garden
  Street, Cambridge, MA, 02138, USA}

\author[0000-0003-3773-5142]{Jason D. Eastman}  % exofast
\affiliation{Center for Astrophysics $|$ Harvard \& Smithsonian, 60 Garden
  Street, Cambridge, MA, 02138, USA}

\author[0000-0001-8812-0565]{Joseph E. Rodriguez}  % exofast
\affiliation{Center for Astrophysics $|$ Harvard \& Smithsonian, 60 Garden
  Street, Cambridge, MA, 02138, USA}

\author{Nicola Astudillo-Defru}  % harps
\affiliation{Departamento de Matem\'atica y F\'isica Aplicadas,
Universidad Cat\'olica de la Sant\'isima Concepci\'on, Alonso de Rivera 2850, 
Concepci\'on, Chile}

\author{Xavier Bonfils}  % harps
\affiliation{CNRS, IPAG, Universit\'e Grenoble Alpes, 38000 Grenoble, France}

\author[0000-0001-7254-4363]{Annelies Mortier}  % harpsn
\affiliation{Astrophysics Group, Cavendish Laboratory, University of Cambridge,
  J.J. Thomson Avenue, Cambridge CB3 0HE, UK}

\author{Christopher A. Watson}  % harpsn observer
\affiliation{Astrophysics Research Centre, School of Mathematics and Physics,
Queen's University Belfast, Belfast, BT7 1NN, UK}

\author{Manu Stalport}  % harpsn observer
\affiliation{Observatoire Astronomique de l'Universit\'e de Gen\`eve, 51 chemin
  des Maillettes, 1290 Versoix, Switzerland}

\author[0000-0002-4445-1845]{Matteo Pinamonti} % harpsn observer
\affiliation{INAF - Osservatorio Astrofisico di Torino, Strada Osservatorio 20,
  Pino Torinese (To) 10025, Italy}

\author[0000-0003-4047-0771]{Florian Lienhard}  % harpsn observer
\affiliation{Astrophysics Group, Cavendish Laboratory, University of Cambridge,
  J.J. Thomson Avenue, Cambridge CB3 0HE, UK}

\author{Avet Harutyunyan} % harpsn observer
\affiliation{Fundaci\'on Galileo Galilei-INAF, Rambla Jos\'e Ana Fernandez
P\'erez 7, 38712 Bre\~{n}a Baja, TF, Spain}

\author{Mario Damasso}  % harpsn
\affiliation{INAF - Osservatorio Astrofisico di Torino, Strada Osservatorio 20,
  Pino Torinese (To) 10025, Italy}

\author{David W. Latham}  % harpsn
\affiliation{Center for Astrophysics $|$ Harvard \& Smithsonian, 60 Garden
  Street, Cambridge, MA, 02138, USA}

\author[0000-0001-6588-9574]{Karen A. Collins}  % lco
\affiliation{Center for Astrophysics $|$ Harvard \& Smithsonian, 60 Garden
  Street, Cambridge, MA, 02138, USA}

\author[0000-0001-8879-7138]{Robert Massey}  % lco
\affiliation{American Association of Variable Star Observers (AAVSO), 49 Bay 
State Rd, Cambridge, MA, 02138, USA}

\author{Jonathan Irwin}  % tres/mearth
\affiliation{Center for Astrophysics $|$ Harvard \& Smithsonian, 60 Garden
  Street, Cambridge, MA, 02138, USA}

\author[0000-0001-6031-9513]{Jennifer G. Winters}  % tres
\affiliation{Center for Astrophysics $|$ Harvard \& Smithsonian, 60 Garden
  Street, Cambridge, MA, 02138, USA}

\author[0000-0002-9003-484X]{David Charbonneau}  % harps/mearth
\affiliation{Center for Astrophysics $|$ Harvard \& Smithsonian, 60 Garden
  Street, Cambridge, MA, 02138, USA}

\author[0000-0002-0619-7639]{Carl Ziegler}  % soar
\affiliation{Dunlap Institute for Astronomy and Astrophysics, University of
  Toronto, 50 St. George Street, Toronto, Ontario M5S 3H4, Canada}  % soar

\author[0000-0003-0593-1560]{Elisabeth Matthews}   % niri
\affiliation{Department of Earth, Atmospheric and Planetary Sciences, and 
Kavli Institute for Astrophysics and Space Research, Massachusetts Institute 
of Technology, Cambridge, MA 02139, USA}

\author{Ian J. M. Crossfield}  % niri
\affiliation{Department of Physics \& Astronomy, University of Kansas, 1082
  Malott, 1251 Wescoe Hall Dr. Lawrence, KS, 66045, USA}

\author{Laura Kreidberg}  % atmosphere calculations
\affiliation{Center for Astrophysics $|$ Harvard \& Smithsonian, 60 Garden
  Street, Cambridge, MA, 02138, USA}

\author[0000-0002-8964-8377]{Samuel N. Quinn} 
\affiliation{Center for Astrophysics $|$ Harvard \& Smithsonian, 60 Garden
  Street, Cambridge, MA, 02138, USA}

\author{George Ricker}
\affiliation{Department of Earth, Atmospheric and Planetary Sciences, and 
Kavli Institute for Astrophysics and Space Research, Massachusetts Institute 
of Technology, Cambridge, MA 02139, USA}

\author[0000-0001-6763-6562]{Roland Vanderspek}
\affiliation{Department of Earth, Atmospheric and Planetary Sciences, and 
Kavli Institute for Astrophysics and Space Research, Massachusetts Institute 
of Technology, Cambridge, MA 02139, USA}

\author[0000-0002-6892-6948]{Sara Seager}
\affiliation{Department of Physics and Kavli Institute for Astrophysics and
Space Research, Massachusetts Institute of Technology, Cambridge, MA 02139, USA}
\affiliation{Department of Earth, Atmospheric and Planetary Sciences,
Massachusetts Institute of Technology, Cambridge, MA 02139, USA}
\affiliation{Department of Aeronautics and Astronautics, MIT, 77 Massachusetts
Avenue, Cambridge, MA 02139, USA}

\author[0000-0002-4265-047X]{Joshua Winn}
\affiliation{Department of Astrophysical Sciences, Princeton University,
Princeton, NJ 08544, USA}

\author{Jon M. Jenkins}
\affiliation{NASA Ames Research Center, Moffett Field, CA, 94035, USA}

\author{Michael Vezie}  % POC
\affiliation{Department of Earth, Atmospheric and Planetary Sciences, and 
Kavli Institute for Astrophysics and Space Research, Massachusetts Institute 
of Technology, Cambridge, MA 02139, USA}

\author{St\'ephane Udry}  % harps/harpsn
\affiliation{Observatoire Astronomique de l'Universit\'e de Gen\`eve, 51 chemin
  des Maillettes, 1290 Versoix, Switzerland}

\author[0000-0002-6778-7552]{Joseph D. Twicken}  % SPOC
\affiliation{NASA Ames Research Center, Moffett Field, CA, 94035, USA}
\affiliation{SETI Institute, Mountain View, CA 94043, USA}

\author[0000-0002-1949-4720]{Peter Tenenbaum}  % SPOC
\affiliation{NASA Ames Research Center, Moffett Field, CA, 94035, USA}

\author[0000-0002-7504-365X]{Alessandro Sozzetti}  % harpsn
\affiliation{INAF - Osservatorio Astrofisico di Torino, Strada Osservatorio 20,
  Pino Torinese (To) 10025, Italy}

\author{Damien S\'egransan}  % harps/harpsn
\affiliation{Observatoire Astronomique de l'Universit\'e de Gen\`eve, 51 chemin
  des Maillettes, 1290 Versoix, Switzerland}

\author[0000-0001-5347-7062]{Joshua E. Schlieder}  % niri
\affiliation{NASA Goddard Space Flight Center, 8800 Greenbelt Rd, Greenbelt,
  MD 20771, USA}

\author[0000-0001-7014-1771]{Dimitar Sasselov}  % harpsn
\affiliation{Center for Astrophysics $|$ Harvard \& Smithsonian, 60 Garden
  Street, Cambridge, MA, 02138, USA}

\author[0000-0003-4422-2919]{Nuno C. Santos}  % harps
\affiliation{Instituto de Astrof\'isica e Ci\^encias do Espa\c{c}o,
  Universidade do Porto, CAUP, Rua das Estrelas, 4150-762 Porto, Portugal}
\affiliation{Departamento de F\'isica e Astronomia, Faculdade de Ci\^encias,
  Universidade do Porto, Rua do Campo Alegre, 4169-007 Porto, Portugal}

\author{Ken Rice}  % harpsn
\affiliation{SUPA, Institute for Astronomy, University of Edinburgh, Blackford
  Hill, Edinburgh, EH9 3HJ, Scotland, UK}

\author[0000-0002-3627-1676]{Benjamin V. Rackham}  % TSO
\affiliation{Department of Earth, Atmospheric and Planetary Sciences, and 
Kavli Institute for Astrophysics and Space Research, Massachusetts Institute 
of Technology, Cambridge, MA 02139, USA}
\affiliation{51 Pegasi b Fellow}

\author[0000-0003-1200-0473]{Ennio Poretti}  % harpsn
\affiliation{Fundaci\'on Galileo Galilei-INAF, Rambla Jos\'e Ana Fernandez
P\'erez 7, 38712 Bre\~{n}a Baja, TF, Spain}
\affiliation{INAF-Osservatorio Astronomico di Brera, via E. Bianchi 46, 23807
Merate (LC), Italy}

\author{Giampaolo Piotto}  % harpsn
\affiliation{Dip. di Fisica e Astronomia Galileo Galilei - Universit\`a di
Padova, Vicolo dell'Osservatorio 2, 35122, Padova, Italy}

\author{David Phillips}  % harpsn
\affiliation{Center for Astrophysics $|$ Harvard \& Smithsonian, 60 Garden
  Street, Cambridge, MA, 02138, USA}

\author{Francesco Pepe}  % harps
\affiliation{Observatoire Astronomique de l'Universit\'e de Gen\`eve, 51 chemin
  des Maillettes, 1290 Versoix, Switzerland}

\author[0000-0002-1742-7735]{Emilio Molinari}  % harpsn
\affiliation{INAF - Osservatorio Astronomico di Cagliari, via della Scienza 5,
09047, Selargius, Italy}

\author{Lucile Mignon}   % harps
\affiliation{CNRS, IPAG, Universit\'e Grenoble Alpes, 38000 Grenoble, France}

\author[0000-0002-9900-4751]{Giuseppina Micela}  % harpsn
\affiliation{INAF - Osservatorio Astronomico di Palermo, Piazza del Parlamento
1, I-90134 Palermo, Italy}

\author{Claudio Melo}   % harps
\affiliation{European Southern Observatory, Alonso de Córdova 3107, Vitacura,
Región Metropolitana, Chile}

\author[0000-0001-8218-1586]{Jos\'e R. de Medeiros}  % harps
\affiliation{Departamento de F\'isica, Universidade Federal do Rio Grande do
  Norte, 59072-970 Natal, RN, Brazil}

\author{Michel Mayor}  % harpsn
\affiliation{Observatoire Astronomique de l'Universit\'e de Gen\`eve, 51 chemin
  des Maillettes, 1290 Versoix, Switzerland}

\author[0000-0001-7233-7508]{Rachel A. Matson}   % niri
\affiliation{U.S. Naval Observatory, Washington, DC 20392, USA}

\author{Aldo F. Martinez Fiorenzano}  % harpsn staff
\affiliation{Fundaci\'on Galileo Galilei-INAF, Rambla Jos\'e Ana Fernandez
P\'erez 7, 38712 Bre\~{n}a Baja, TF, Spain}

\author[0000-0003-3654-1602]{Andrew W. Mann}  % soar
\affiliation{Department of Physics and Astronomy, The University of North
  Carolina at Chapel Hill, Chapel Hill, NC 27599-3255, USA}

\author[0000-0003-1259-4371]{Antonio Magazz\'u}  % harpsn staff
\affiliation{Fundaci\'on Galileo Galilei-INAF, Rambla Jos\'e Ana Fernandez
P\'erez 7, 38712 Bre\~{n}a Baja, TF, Spain}

\author{Christophe Lovis}  % harps
\affiliation{Observatoire Astronomique de l'Universit\'e de Gen\`eve, 51 chemin
  des Maillettes, 1290 Versoix, Switzerland}

\author[0000-0003-3204-8183]{Mercedes L\'opez-Morales}  % harpsn
\affiliation{Center for Astrophysics $|$ Harvard \& Smithsonian, 60 Garden
  Street, Cambridge, MA, 02138, USA}

\author{Eric Lopez}   % harpsn
\affiliation{NASA Goddard Space Flight Center, 8800 Greenbelt Rd, Greenbelt, MD
  20771, USA}

\author[0000-0001-6513-1659]{Jack J. Lissauer}  % radval discussion
\affiliation{NASA Ames Research Center, Moffett Field, CA, 94035, USA}

\author{S\'ebastien L\'epine}  % TSO
\affiliation{Department of Physics and Astronomy, Georgia State University,
  Atlanta, GA 30302, USA}

\author{Nicholas Law}  % soar
\affiliation{Department of Physics and Astronomy, The University of North
  Carolina at Chapel Hill, Chapel Hill, NC 27599-3255, USA}

\author[0000-0003-0497-2651]{John F. Kielkopf}  % lco general
\affiliation{Department of Physics and Astronomy, University of Louisville,
Louisville, KY 40292, USA}

\author{John A. Johnson}  % harpsn
\affiliation{Center for Astrophysics $|$ Harvard \& Smithsonian, 60 Garden
  Street, Cambridge, MA, 02138, USA}

\author[0000-0002-4625-7333]{Eric L. N. Jensen}  % lco
\affiliation{Dept. of Physics \& Astronomy, Swarthmore College, Swarthmore PA
  19081, USA}

\author[0000-0002-2532-2853]{Steve B. Howell}  % niri
\affiliation{NASA Ames Research Center, Moffett Field, CA, 94035, USA}

\author{Erica Gonzales}  % niri
\affiliation{Department of Astronomy and Astrophysics, University
of California, Santa Cruz, CA 95064, USA}

\author{Adriano Ghedina}  % harpsn staff
\affiliation{Fundaci\'on Galileo Galilei-INAF, Rambla Jos\'e Ana Fernandez
P\'erez 7, 38712 Bre\~{n}a Baja, TF, Spain}

\author[0000-0003-0536-4607]{Thierry Forveille}  % harps
\affiliation{CNRS, IPAG, Universit\'e Grenoble Alpes, 38000 Grenoble, France}

\author[0000-0001-8504-283X]{Pedro Figueira}  % harps
\affiliation{European Southern Observatory, Alonso de C\'ordova 3107, Vitacura,
Regi\'on Metropolitana, Chile}
\affiliation{Instituto de Astrof\'isica e Ci\^encias do Espa\c{c}o,
  Universidade do Porto, CAUP, Rua das Estrelas, 4150-762 Porto, Portugal}

\author{Xavier Dumusque}  % harpsn
\affiliation{Observatoire Astronomique de l'Universit\'e de Gen\`eve, 51 chemin
  des Maillettes, 1290 Versoix, Switzerland}

\author[0000-0001-8189-0233]{Courtney D. Dressing}  % TSO
\affiliation{Astronomy Department, University of California, Berkeley, CA,
  94720, USA}

\author{Ren\'e Doyon}  % harps
\affiliation{D\'epartement de physique, Universit\'e de Montr\'eal, 2900 boul.
  \'Edouard-Montpetit, Montr\'eal, QC, H3C 3J7, Canada}

\author[0000-0001-9289-5160]{Rodrigo F. D\'iaz}  % harps
\affiliation{International Center for Advanced Studies (ICAS) and ICIFI 
(CONICET), ECyT-UNSAM, Campus Miguelete, 25 de Mayo y Francia, (1650) 
Buenos Aires, Argentina}

\author{Luca Di Fabrizio}  % harpsn staff
\affiliation{Fundaci\'on Galileo Galilei-INAF, Rambla Jos\'e Ana Fernandez
P\'erez 7, 38712 Bre\~{n}a Baja, TF, Spain}

\author{Xavier Delfosse}  % harps
\affiliation{CNRS, IPAG, Universit\'e Grenoble Alpes, 38000 Grenoble, France}

\author{Rosario Cosentino}  % harpsn staff
\affiliation{Fundaci\'on Galileo Galilei-INAF, Rambla Jos\'e Ana Fernandez
P\'erez 7, 38712 Bre\~{n}a Baja, TF, Spain}

\author[0000-0003-2239-0567]{Dennis M. Conti}  % lco general
\affiliation{American Association of Variable Star Observers, 49 Bay State
Road, Cambridge, MA 02138, USA}

\author[0000-0003-2781-3207]{Kevin I. Collins}  % lco general
\affiliation{George Mason University, 4400 University Drive, Fairfax, VA,
22030 USA}

\author{Andrew Collier Cameron}  % harpsn
\affiliation{School of Physics and Astronomy, University of St Andrews, North
Haugh, St Andrews, Fife, KY16 9SS, UK}

\author{David Ciardi}  % niri
\affiliation{Caltech/IPAC, 1200 E. California Blvd. Pasadena, CA 91125, USA}

\author[0000-0003-1963-9616]{Douglas A. Caldwell}  % SPOC
\affiliation{NASA Ames Research Center, Moffett Field, CA, 94035, USA}

\author{Christopher Burke}  % POC
\affiliation{Department of Earth, Atmospheric and Planetary Sciences, and 
Kavli Institute for Astrophysics and Space Research, Massachusetts Institute 
of Technology, Cambridge, MA 02139, USA}

\author{Lars Buchhave}  % harpsn
\affiliation{DTU Space, National Space Institute, Technical University of 
Denmark, Elektrovej 328, DK-2800 Kgs. Lyngby, Denmark}

\author{C\'esar Brice\~{n}o}  % soar
\affiliation{Cerro Tololo Inter-American Observatory, Casilla 603, La Serena,
  Chile}

\author{Patricia Boyd}  % POC
\affiliation{NASA Goddard Space Flight Center, 8800 Greenbelt Rd, Greenbelt, MD
  20771, USA}

\author{Fran\c{c}ois Bouchy}  % harps
\affiliation{Observatoire Astronomique de l'Universit\'e de Gen\`eve, 51 chemin
  des Maillettes, 1290 Versoix, Switzerland}

\author{Charles Beichman}  % niri
\affiliation{NASA Exoplanet Science Institute, Infrared Processing \& Analysis
  Center, Jet Propulsion Laboratory, California Institute of Technology,
  Pasadena CA, 91125, USA}

\author{\'Etienne Artigau}  % harps
\affiliation{D\'epartement de physique, Universit\'e de Montr\'eal, 2900 boul.
  \'Edouard-Montpetit, Montr\'eal, QC, H3C 3J7, Canada}

\author{Jose M. Almenara}  % harps
\affiliation{CNRS, IPAG, Universit\'e Grenoble Alpes, 38000 Grenoble, France}

%% file: ltt3780table.tex
%\capstartfalse
\begin{deluxetable}{lcc}
\tabletypesize{\small}
\tablecaption{LTT 3780 stellar parameters.\label{tab:star}}
\tablewidth{0pt}
\tablehead{\colhead{Parameter} & \colhead{Value} & \colhead{Refs}}
%\decimalcolnumbers
\startdata 
%\vspace{0.05pt} \\
\multicolumn{3}{c}{\emph{LTT 3780, LP 729-54, TIC 36724087, TOI-732}} \\
%\vspace{1pt} \\
\multicolumn{3}{c}{\emph{Astrometry}} \\
Right ascension (J2000.0), $\alpha$ & 10:18:34.78 & 1,2 \\
Declination (J2000.0), $\delta$ & -11:43:04.08 & 1,2 \\
RA proper motion, $\mu_{\alpha}$ [mas yr$^{-1}$] & $-341.41\pm 0.11$ & 1,2 \\
Dec proper motion, $\mu_{\delta}$ [mas yr$^{-1}$] & $-247.87\pm 0.11$ & 1,2 \\
Parallax, $\varpi$ [mas] & $45.493\pm 0.083$ & 1,2 \\
Distance, $d$ [pc] & $21.981\pm 0.040$ & 1,2 \\
\multicolumn{3}{c}{\emph{Photometry}} \\
%$B$ & $\pm  & 3 \\
$V$ & $13.07\pm 0.015$ & 3 \\
%$R$ & $11.94$ & ? \\
%$I$ & $10.44$ & ? \\
$G_{\text{BP}}$ & $13.352\pm 0.004$ & 1,4 \\
$G$ & $11.8465\pm 0.0005$ & 1,4 \\
$G_{\text{RP}}$ & $10.658\pm 0.002$ & 1,4 \\
$T$ & $10.585\pm 0.007$ & 5 \\
$J$ & $9.007\pm 0.030$ & 6 \\
$H$ & $8.439\pm 0.065$ & 6 \\
$K_s$ & $8.204\pm 0.021$ & 6 \\
$W_1$ & $8.037 \pm 0.022$ & 7 \\
$W_2$ & $7.880 \pm 0.019$ & 7 \\
$W_3$ & $7.771 \pm 0.019$ & 7 \\
$W_4$ & $7.577 \pm 0.166$ & 7 \\
\multicolumn{3}{c}{\emph{Stellar parameters}} \\
Spectral type & M4V & 8 \\
$M_V$ & $11.36\pm 0.02$ & 9 \\ 
$M_{K_s}$ & $6.49\pm 0.02$ & 9 \\ 
Effective temperature, \teff{} [K] & $3331\pm 157$ & 5 \\
Surface gravity, \logg{} [dex] & $4.896\pm 0.029$ & 9 \\
Metallicity, [Fe/H] [dex] & $0.28^{+0.11}_{-0.13}$ & 9 \\
Stellar radius, $R_s$ [R$_{\odot}$] & $0.374\pm 0.011$ & 9 \\ 
Stellar mass, $M_s$ [M$_{\odot}$] & $0.401\pm 0.012$ & 9 \\
\vspace{-0.15cm} Projected rotation velocity, && \\ \vspace{-0.25cm}
& $<1.3$ & 9 \\
\vsini{} [km s$^{-1}$] && \\
$\log{R_{\text{HK}}'}$ & $-5.59\pm 0.09$ & 9 \\
\vspace{-0.15cm} Estimated rotation period, && \\ \vspace{-0.25cm}
& $104\pm 15$ & 9 \\
\prot{} [days] && \\
\enddata
\tablecomments{\textbf{References:}
  1) \citealt{gaia18}
  2) \citealt{lindegren18}
  3) \citealt{reid02}
  4) \citealt{evans18}
  5) \citealt{stassun19}
  6) \citealt{cutri03}
  7) \citealt{cutri13}
  8) \citealt{scholz05}
  9) this work.}
\end{deluxetable}
%\capstarttrue

%% file: rvtable.tex
\begin{deluxetable}{cccc}
\tablecaption{Radial velocity time series of LTT 3780 from HARPS \& HARPS-N\label{tab:rvs}}
\tablewidth{0pt}
\tablehead{Time & RV & $\sigma_{\text{RV}}$ & Instrument \\
$[$BJD - 2,457,000$]$ & $[\text{m s}^{-1}]$ & $[\text{m s}^{-1}]$ & }
\startdata
%1656.493768 &     4.365 &   2.070 &    HARPS \\
%1661.467140 &    -3.240 &  1.404 &    HARPS \\
%1802.851053 &    2.572 &  1.023 &    HARPS \\
%1803.846779 &   -5.597 &  1.108 &    HARPS \\
%1805.845559 &    -2.572 &  1.921 &    HARPS \\
%1806.845616 &    -9.207 &  1.204 &    HARPS \\
%1807.846332 &    -5.480 &  1.253 &    HARPS \\
1821.837965 &   -0.959 &   1.310 &    HARPS \\
1831.760260 &   -10.056 &  1.330 &  HARPS-N \\
1836.858657 &    -5.946 &   1.403 &    HARPS \\
\enddata
%\tablecomments{For conciseness, only a subset of three rows are depicted here to illustrate the table's contents. The entirety of this table is provided in the arXiv source code.}
\end{deluxetable}

%% file: priors_exo.tex
\begin{deluxetable*}{lcc}
\tabletypesize{\small}
\tablecaption{\tess{} light curve and RV model parameter priors\label{tab:priors}}
\tablewidth{0pt}
\tablehead{Parameter & Fiducial Model Priors & \texttt{EXOFASTv2} Model Priors}
\startdata
\multicolumn{3}{c}{\emph{Stellar parameters}} \\
\teff{,} [K] & $\mathcal{N}(3331,157)$ & $\mathcal{N}(3351,150)$ \\
$M_s$, [\Msun{]} & $\mathcal{N}(0.401,0.012)$ & $\mathcal{N}(0.401,0.012)$ \\
$R_s$, [\Rsun{]} & $\mathcal{N}(0.374,0.011)$ & $\mathcal{N}(0.374,0.011)$ \\
\multicolumn{3}{c}{\emph{Light curve hyperparameters}} \\
$f_0$ & $\mathcal{N}(0,10)$ & $\mathcal{U}(-\inf,\inf)$ \\ 
$\ln{\omega_0}$, [days$^{-1}$] & $\mathcal{N}(0,10)$ & - \\
$\ln{S_0\omega_0^4}$ & $\mathcal{N}(\ln{\text{var}(\mathbf{f}_{\text{TESS}})},10)$ & - \\
$\ln{s_{\text{TESS}}^2}$ & $\mathcal{N}(\ln{\text{var}(\mathbf{f}_{\text{TESS}})},10)$ & - \\
$u_1$ & $\mathcal{U}(0,1)$ & $\mathcal{U}(0.225,0.425)$ \\
$u_2$ & $\mathcal{U}(0,1)$ & $\mathcal{U}(0.232,0.432)$ \\
Dilution & - & $\mathcal{N}(0,0.1\: \delta)$\tablenotemark{a} \\ 
\multicolumn{3}{c}{\emph{RV parameters}} \\
$\ln{\lambda}$, [days] & $\mathcal{U}(\ln{1},\ln{1000})$ & - \\
$\ln{\Gamma}$ & $\mathcal{U}(-3,3)$ & -  \\
$\ln{P_{\text{GP}}}$, [days] & $\mathcal{N}(\ln{104},\ln{30})$\tablenotemark{b} & -  \\
$\ln{a}_{\text{HARPS}}$, [\mps{]} & $\mathcal{U}(-5,5)$ & -  \\
$\ln{a}_{\text{HARPS-N}}$, [\mps{]} & $\mathcal{U}(-5,5)$ & -  \\
$\ln{s}_{\text{HARPS}}$, [\mps{]} & $\mathcal{U}(-5,5)$ & $\mathcal{U}(-\inf,\inf)$ \\
$\ln{s}_{\text{HARPS-N}}$, [\mps{]} & $\mathcal{U}(-5,5)$ & $\mathcal{U}(-\inf,\inf)$ \\
$\gamma_{\text{HARPS}}$, [\mps{]} & $\mathcal{U}(-185,205)$ & $\mathcal{U}(-\inf,\inf)$ \\
$\gamma_{\text{HARPS-N}}$, [\mps{]} & $\mathcal{U}(-185,205)$ & $\mathcal{U}(-\inf,\inf)$ \\
\multicolumn{3}{c}{\emph{LTT 3780b parameters}} \\
$\ln{P}_b$, [days] & $\mathcal{N}(\ln{0.768},0.5)$ & - \\
$P_b$, [days] & - & $\mathcal{U}(-\inf,\inf)$ \\
$T_{0,b}$, [BJD-2,457,000] & $\mathcal{N}(1543.911,0.5)$ & $\mathcal{U}(1543.7,1544.2)$ \\
$\ln{r_{p,b}}$, [\Rearth{]} & $\mathcal{N}(0.5\cdot \ln(Z_b) + \ln{R_s},1)$\tablenotemark{c} & -\\
$r_{p,b}/R_s$ & - & $\mathcal{U}(-\inf,\inf)$ \\
$b_b$ & $\mathcal{U}(0,1+r_{p,b}/R_s)$ & - \\
$\ln{K}_b$, [\mps{]} & $\mathcal{U}(-5,5)$ & - \\
$K_b$, [\mps{]} & - & $\mathcal{U}(-\inf,\inf)$ \\
\multicolumn{3}{c}{\emph{LTT 3780c parameters}} \\
$\ln{P}_c$, [days] & $\mathcal{N}(\ln{12.254},0.5)$ & - \\
$P_c$, [days] & - & $\mathcal{U}(-\inf,\inf)$ \\
$T_{0,c}$, [BJD-2,457,000] & $\mathcal{N}(1546.848,0.5)$ & $\mathcal{U}(1542.8,1550.9)$ \\
$\ln{r_{p,c}}$, [\Rearth{]} & $\mathcal{N}(0.5\cdot \ln(Z_c) + \ln{R_s},1)$\tablenotemark{d} \\
$r_{p,c}/R_s$ & - & $\mathcal{U}(-\inf,\inf)$ \\
$b_c$ & $\mathcal{U}(0,1+r_{p,c}/R_s)$ \\
$\ln{K}_c$, [\mps{]} & $\mathcal{U}(-5,5)$ & - \\
$K_c$, [\mps{]} & - & $\mathcal{U}(-\inf,\inf)$ \\
$e_c$ & $\mathcal{B}(0.867,3.03)$\tablenotemark{e} \\
$\omega_c$, [rad] & $\mathcal{U}(-\pi,\pi)$ \\
\enddata
\tablecomments{Gaussian distributions are denoted by $\mathcal{N}$ and are
  parameterized by mean and standard deviation values. Uniform distributions
  are denoted by $\mathcal{U}$ and bounded by the specified lower and upper
  limits. Beta distributions are denoted by $\mathcal{B}$ and are parameterized
  by the shape parameters $\alpha$ and $\beta$.}
\tablenotetext{a}{$delta$ is the SPOC-derived dilution factor applied to the \tess{} light curve.}
\tablenotetext{b}{$P_{\text{GP}}$ is constrained by the estimate of the stellar rotation period
from $\log{R_{\text{HK}}'}$ whose uncertainty is artificially inflated.}
\tablenotetext{c}{The transit depth of TOI-732.01 reported by the SPOC: $Z_b=1253$ ppm.}
\tablenotetext{d}{The transit depth of TOI-732.02 reported by the SPOC: $Z_b=3417$ ppm.}
\tablenotetext{e}{\citealt{kipping13}.}
\end{deluxetable*}

%% file: tsmtable.tex
\begin{deluxetable*}{lcccccccccccc}
\tabletypesize{\small}
\tablecaption{Transmission spectroscopy metric values for warm mini-Neptunes\tablenotemark{a}\label{tab:tsm}}
\tablewidth{0pt}
\tablehead{Planet &      $P$ &     $r_p$    & $m_p$  &  $Z$ &      $T_{\text{eq}}$\tablenotemark{b} &    $J$ &    \teff{} &     $R_s$ &     $M_s$ &       TSM & TSM- & Refs \\
name & [days] & [\Rearth{]} & [\Mearth{]}  & [ppt] & [K] & mag & [K] & [\Rsun{]} & [\Msun{]} & & normalized & }
\startdata
TOI-270d    &  11.38 &  2.13 &   5.48\tablenotemark{c} &  2.6 &  372 &  9.099 &  3386 &  0.38 &  0.40 &  86.8 & 1.00 & 1 \\
TOI-700c    &  16.05 &  2.63 &   7.64\tablenotemark{c} &  3.3 &  356 &  9.469 &  3480 &  0.42 &  0.42 &  77.5 & 0.89 & 2,3 \\
\textbf{LTT 3780c}   &  12.25 &  2.30 &   8.59 &  3.3 &  353 &  9.007 &  3331 &  0.37 &  0.40 &  71.5 & 0.82 & 4 \\
HD 15337c   &  17.17 &  2.52 &   8.79 &  0.6 &  648 &  7.553 &  5125 &  0.87 &  0.90 &  60.6 & 0.70 &  5 \\
GJ 143b     &  35.61 &  2.61 &  22.70 &  1.2 &  427 &  6.081 &  4640 &  0.70 &  0.73 &  53.0 & 0.61 & 6 \\
K2-266d     &  14.70 &  2.93 &   8.90 &  1.5 &  538 &  9.611 &  4285 &  0.70 &  0.69 &  47.1 & 0.54 & 7 \\
K2-18b      &  32.94 &  2.71 &   8.63 &  2.8 &  290 &  9.763 &  3505 &  0.47 &  0.50 &  42.8 & 0.49 & 8 \\
Kepler-96b  &  15.24 &  2.67 &   8.46 &  0.6 &  798 &  9.260 &  5690 &  1.02 &  1.00 &  30.6 & 0.35 & 9 \\
K2-266e     &  19.48 &  2.73 &  14.30 &  1.3 &  490 &  9.611 &  4285 &  0.70 &  0.69 &  21.3 & 0.24 & 7 \\
Kepler-102e &  16.15 &  2.22 &   8.93 &  0.7 &  604 &  9.984 &  4909 &  0.76 &  0.81 &  16.3 & 0.19 & 9 \\
HD 119130b  &  16.98 &  2.63 &  24.50 &  0.5 &  801 &  8.730 &  5725 &  1.09 &  1.00 &  11.3 & 0.13 & 10 \\
K2-38c      &  10.56 &  2.42 &   9.90 &  0.3 &  928 &  9.911 &  5757 &  1.38 &  2.24 &   9.2 & 0.11 & 11 \\
\enddata
\tablecomments{\textbf{References:}
  1) \citealt{gunther19}
  2) \citealt{gilbert20}
  3) \citealt{rodriguez20}
  4) this work
  5) \citealt{dumusque19}
  6) \citealt{dragomir19}
  7) \citealt{rodriguez18}
  8) \citealt{cloutier19a}
  9) \citealt{marcy14}
  10) \citealt{luque19b}
  11) \citealt{sinukoff16}.
}
\tablenotetext{a}{Here we define warm mini-Neptunes as having $P\in [10,40]$ days and $r_p \in [2,3]$ \Rearth{.}}
\tablenotetext{b}{$T_{\text{eq}}$ is calculated assuming zero albedo and full heat redistribution.}
\tablenotetext{c}{Planet masses are estimated using the mass-radius relation implemented in the \texttt{forecaster} code \citep{chen17}.}
\end{deluxetable*}

%% file: esmtable.tex
\begin{deluxetable*}{lcccccccccccc}
\tabletypesize{\small}
\tablecaption{Emission spectroscopy metric values for select close-in Earth-sized planets\tablenotemark{a}\label{tab:esm}}
\tablewidth{0pt}
\tablehead{Planet &      $P$ &     $r_p$ &    $Z$ &      $T_{\text{eq}}$\tablenotemark{b} &         $T_{\text{day}}$\tablenotemark{c} &    $K_s$ &    \teff{} &     $R_s$ &     $M_s$ &       ESM & ESM- & Refs \\
name & [days] & [\Rearth{]} & [ppt] & [K] & [K] & mag & [K] & [\Rsun{]} & [\Msun{]} & & normalized}
\startdata
LHS 3844b   &  0.46 &  1.30 &  4.0 &   805 &   886 &   9.145 &  3036 &  0.19 &  0.15 &  29.0 & 1.00 & 1 \\
GJ 1252b    &  0.52 &  1.19 &  0.8 &  1089 &  1198 &   7.915 &  3458 &  0.39 &  0.38 &  16.4 & 0.57 & 2 \\
\textbf{LTT 3780b}   &  0.77 &  1.33 &  1.1 &   892 &   982 &   8.204 &  3331 &  0.37 &  0.40 &  13.4 & 0.46 & 3 \\
L 168-9b    &  1.40 &  1.39 &  0.5 &   981 &  1079 &   7.082 &  3800 &  0.60 &  0.62 &   9.9 & 0.34 & 4 \\
GJ 1132b    &  1.63 &  1.13 &  2.4 &   585 &   643 &   8.322 &  3270 &  0.21 &  0.18 &   9.5 & 0.33 & 5 \\
L 98-59c    &  3.69 &  1.35 &  1.6 &   515 &   566 &   7.101 &  3412 &  0.31 &  0.31 &   6.9 & 0.24 & 6 \\
LTT 1445Ab  &  5.36 &  1.38 &  2.0 &   435 &   478 &   6.500 &  3335 &  0.28 &  0.26 &   6.3 & 0.22 & 7 \\
LP 791-18b  &  0.95 &  1.12 &  3.6 &   594 &   653 &  10.644 &  2949 &  0.17 &  0.14 &   5.9 & 0.20 & 8 \\
L 98-59b    &  2.25 &  0.80 &  0.6 &   607 &   668 &   7.101 &  3412 &  0.31 &  0.31 &   4.1 & 0.14 & 6 \\
TRAPPIST-1b &  1.51 &  1.09 &  6.9 &   405 &   446 &  10.300 &  2559 &  0.12 &  0.08 &   4.0 & 0.14 & 9 \\
LHS 1140c   &  3.78 &  1.28 &  3.1 &   434 &   477 &   8.821 &  3216 &  0.21 &  0.18 &   3.4 & 0.12 & 10 \\
\enddata
\tablecomments{\textbf{References:}
  1) \citealt{vanderspek19}
  2) \citealt{shporer20}
  3) this work
  4) \citealt{astudillodefru20}
  5) \citealt{berta15}
  6) \citealt{kostov19}
  7) \citealt{winters19b}
  8) \citealt{crossfield19}
  9) \citealt{gillon17a}
  10) \citealt{ment19}  
}
\tablenotetext{a}{Here we define Earth-sized planets as those with $r_p < 1.5$ \Rearth{.}}
\tablenotetext{b}{$T_{\text{eq}}$ is calulated assuming zero albedo and full heat redistribution.}
\tablenotetext{c}{For the purpose of calculating ESM values, we assume that $T_{\text{day}} = 1.1T_{\text{eq}}$ for all planets.}
\end{deluxetable*}

%% file: tesstable_exofast.tex
\startlongtable
\begin{deluxetable*}{lcc}
\tabletypesize{\footnotesize}
\tablecaption{Point estimates of the LTT 3780 planetary system model parameters\label{tab:tess}}
\tablewidth{0pt}
\tablehead{Parameter & Fiducial Model Values\tablenotemark{a} & \texttt{EXOFASTv2} Model Values\tablenotemark{b}}
\startdata
\multicolumn{3}{c}{\emph{TESS light curve parameters}} \\
Baseline flux, $f_0$ & $1.000072\pm 0.000070$ & $1.000043\pm 0.000038$  \\
$\ln{\omega_0}$ & $1.64\pm 1.15$ & - \\
$\ln{S_0 \omega_0^4}$ & $3.62^{+0.40}_{-0.39}$ & - \\
$\ln{s_{\text{TESS}}^2}$ & $1.21\pm 0.01$ & - \\
\tess{} limb darkening coefficient, $u_1$ & $0.28^{+0.33}_{-0.20}$ & $0.30^{+0.07}_{-0.05}$ \\
\tess{} limb darkening coefficient, $u_2$ & $0.16^{+0.37}_{-0.28}$ & $0.32^{+0.07}_{-0.06}$ \\
Dilution & - & $0.023^{+0.047}_{-0.048}$ \\
\smallskip \\
\multicolumn{3}{c}{\emph{RV parameters}} \\
$\ln{\lambda/\text{day}}$ & $4.5^{+1.0}_{-0.4}$ & - \\
$\ln{\Gamma}$ & $-0.1^{+1.3}_{-1.2}$ & - \\
$\ln{P_{\text{GP}}/\text{day}}$ & $4.64^{+0.14}_{-0.16}$ & - \\
$\ln{a_{\text{HARPS}}/\text{m/s}}$ & $0.52^{+0.69}_{-0.62}$ & - \\
$\ln{a_{\text{HARPS-N}}/\text{m/s}}$ & $1.25^{+0.70}_{-0.74}$ & - \\
Jitter, $s_{\text{HARPS}}$ [\mps{]} & $0.11^{+0.48}_{-0.09}$ & $1.41^{+0.70}_{-0.80}$ \\
Jitter, $s_{\text{HARPS-N}}$ [\mps{]} & $1.24^{+0.36}_{-0.46}$ & $3.54^{+0.99}_{-0.75}$ \\
Systemic velocity, $\gamma_{\text{HARPS}}$ [\mps{]} & $195.5^{+1.4}_{-1.5}$ & $195.4^{+0.5}_{-0.5}$ \\
Systemic velocity, $\gamma_{\text{HARPS-N}}$ [\mps{]} & $196.8^{+4.6}_{-3.6}$ & $194.3^{+1.0}_{-1.0}$ \\
\smallskip \\
\multicolumn{3}{c}{\emph{LTT 3780b (TOI-732.01) parameters}} \\
Log orbital period, $\ln{P_b}$ & $-0.26338\pm 0.00007$ & -  \\
Orbital period, $P_b$ [days] & $0.768448^{+0.000055}_{-0.000053}$ &  $0.7683881^{+0.0000084}_{-0.0000083}$ \\
Time of mid-transit, $T_{0,b}$ [BJD - 2,457,000] & $1543.9115\pm 0.0011$ & $1543.91199^{+0.00059}_{-0.00051}$ \\
Transit duration $D_b$ [hrs] & $0.805^{+0.049}_{-0.072}$ & $0.786^{+0.024}_{-0.020}$ \\
Transit depth, $Z_b$ [ppt] & $1.087^{+0.098}_{-0.103}$ & $1.076^{+0.093}_{-0.089}$ \\
Scaled semimajor axis, $a_b/R_s$ & $7.03^{+0.23}_{-0.21}$ & $7.05^{+0.24}_{-0.22}$ \\
Planet-to-star radius ratio, $r_{p,b}/R_s$ & $0.0330^{+0.0014}_{-0.0016}$ & $0.0328^{+0.0014}_{-0.0014}$ \\
Impact parameter, $b_b$ & $0.35^{+0.20}_{-0.23}$ & $0.43^{+0.08}_{-0.12}$ \\
Inclination, $i_b$ [deg] & $87.1^{+1.8}_{-1.7}$ & $86.5^{+1.0}_{-0.7}$ \\
Eccentricity, $e_b$ & 0 (fixed) & 0 (fixed) \\
Planet radius, $r_{p,b}$ [\Rearth{]} & $1.332^{+0.072}_{-0.075}$ & $1.321^{+0.074}_{-0.073}$ \\
Log RV semi-amplitude, $\ln{K_b}$ & $1.23^{+0.14}_{-0.17}$ & $1.26^{+0.14}_{-0.17}$ \\
RV semi-amplitude, $K_b$ [\mps{]} & $3.41^{+0.63}_{-0.63}$ & $3.54^{+0.54}_{-0.55}$ \\
Planet mass, $m_{p,b}$ [\Mearth{]} & $2.62^{+0.48}_{-0.46}$ & $2.77^{+0.43}_{-0.43}$ \\
Bulk density, $\rho_b$ [g cm$^{-3}$] & $6.1^{+1.8}_{-1.5}$ & $6.5^{+1.7}_{-1.4}$ \\
Surface gravity, $g_b$ [m s$^{-2}$] & $14.4^{+3.7}_{-3.3}$ & $15.5^{+3.6}_{-3.4}$ \\
Escape velocity, $v_{\text{esc},b}$ [km s$^{-1}$] & $15.7^{+1.5}_{-1.5}$ & $16.2^{+1.3}_{-1.4}$ \\
Semimajor axis, $a_b$ [AU] & $0.01211^{+0.00012}_{-0.00012}$ & $0.01212^{+0.00012}_{-0.00012}$ \\
Insolation, $F_b$ [F$_{\oplus}$] & $106^{+22}_{-19}$ & $106^{+23}_{-19}$ \\
Equilibrium temperature, $T_{\text{eq},b}$ [K] && \\
\hspace{2pt} Bond albedo = 0.0 & $892\pm 44$ & $892\pm 44$ \\
\hspace{2pt} Bond albedo = 0.3 & $816\pm 40$ & $816\pm 40$ \\
\smallskip \\
\multicolumn{3}{c}{\emph{LTT 3780c (TOI-732.02) parameters}} \\
Log orbital period, $\ln{P_c}$ & $2.50582\pm 0.00023$ & - \\
Orbital period, $P_c$ [days] & $12.2519^{+0.0028}_{-0.0030}$ & $12.252048^{+0.000060}_{-0.000059}$ \\
Time of mid-transit, $T_{0,c}$ [BJD - 2,457,000] & $1546.8484\pm 0.0014$ & $1546.8481^{+0.0011}_{-0.0012}$ \\
Transit duration $D_c$ [hrs] & $1.392^{+0.050}_{-0.049}$ & $1.404^{+0.048}_{-0.046}$ \\
Transit depth, $Z_c$ [ppt] & $3.24^{+0.41}_{-0.37}$ & $3.13^{+0.28}_{-0.28}$ \\
Scaled semimajor axis, $a_c/R_s$ & $44.6^{+1.5}_{-1.3}$ & $44.7^{+1.5}_{-1.4}$ \\
Planet-to-star radius ratio, $r_{p,c}/R_s$ & $0.0570^{+0.0035}_{-0.0033}$ & $0.0560^{+0.0024}_{-0.0025}$ \\
Impact parameter, $b_c$ & $0.65^{+0.15}_{-0.36}$ & $0.71^{+0.08}_{-0.15}$ \\
Inclination, $i_c$ [deg] & $89.18^{+0.47}_{-0.22}$ & $88.95^{+0.10}_{-0.09}$ \\
$e_c\cos{\omega_c}$ & - & $-0.05^{+0.07}_{-0.08}$ \\
$e_c\sin{\omega_c}$ & - & $0.15^{+0.15}_{-0.13}$ \\
$\sqrt{e_c}\cos{\omega_c}$ & $0.13^{+0.12}_{-0.15}$ & - \\
$\sqrt{e_c}\sin{\omega_c}$ & $0.07^{+0.17}_{-0.19}$ & - \\
Eccentricity, $e_c$ & $0.06^{+0.15}_{-0.14}$ & $0.18^{+0.14}_{-0.11}$ \\
Argument of periastron, $\omega_c$ [deg] & $124^{+87}_{-147}$ & $111^{+39}_{-27}$ \\
Planet radius, $r_{p,c}$ [\Rearth{]} & $2.30^{+0.16}_{-0.15}$ & $2.25^{+0.13}_{-0.13}$ \\
Log RV semi-amplitude, $\ln{K_c}$ & $1.49^{+0.17}_{-0.17}$ & $1.60^{+0.13}_{-0.15}$ \\
RV semi-amplitude, $K_c$ [\mps{]} & $4.44^{+0.82}_{-0.68}$ & $4.94^{+0.68}_{-0.67}$ \\
Planet mass, $m_{p,c}$ [\Mearth{]} & $8.6^{+1.6}_{-1.3}$ & $9.5^{+1.3}_{-1.3}$ \\
Bulk density, $\rho_c$ [g cm$^{-3}$] & $3.9^{+1.0}_{-0.9}$ & $4.6^{+1.1}_{-0.9}$ \\
Surface gravity, $g_c$ [m s$^{-2}$] & $16.0^{+3.7}_{-3.3}$ & $18.3^{+3.5}_{-3.1}$\\
Escape velocity, $v_{\text{esc},c}$ [km s$^{-1}$] & $21.7^{+2.1}_{-2.0}$ & $23.0^{+1.7}_{-1.7}$ \\
Semimajor axis, $a_c$ [AU] & $0.07673^{+0.00075}_{-0.00077}$ & $0.07678^{+0.00076}_{-0.00077}$ \\
Insolation, $F_c$ [F$_{\oplus}$] & $2.63^{+0.56}_{-0.48}$ & $2.63^{+0.56}_{-0.48}$ \\
Equilibrium temperature, $T_{\text{eq},c}$ [K] && \\
\hspace{2pt} Bond albedo = 0.0 & $353\pm 18$ & $354\pm 18$ \\
\hspace{2pt} Bond albedo = 0.3 & $323\pm 16$ & $324\pm 16$ \\
\enddata
\tablenotetext{a}{Our fiducial model features sequential modeling of the \tess{} light curve, with a SHO GP detrending component plus two transiting planets, followed by the RV analysis conditioned on the results of the transit analysis. The fiducial RV model includes a quasi-periodic activity model plus two keplerian orbital solutions. The LTT 3780b keplerian component is fixed to a circular orbit.}
\tablenotetext{b}{Our alternative analysis is a global model of the \tess{} light curve, ground-based light curves, and RVs using the \texttt{EXOFASTv2} software. The input light curves have already been detrended and the residual RV noise is treated as an additive scalar jitter. This global model produces self-consistent results between the transit and RV dataset and improves the precision on each planet's orbital ephemeris by including the ground-based transit light curves.}
%\tablecomments{}
\end{deluxetable*}

%% file: appendices.tex
\appendix
\section{Limits on the planet masses for consistency with models of
photoevaporation} \label{app:pe}
Here we present the formalism used to estimate mass limits on planets spanning
the radius valley within a multi-transiting system under the photoevaporation
model \citep{owen20}.
This model is adopted from \cite{owen17} in which a population of
non-rocky planets is formed with a distribution of Earth-like core masses plus
H/He envelopes. The energy-limited atmospheric mass loss rate due to XUV heating
by the host star, and subsequent thermal escape, is
$\dot{M}_{\text{atm}}=\eta_p \pi r_{\text{core}}^3 L_{\text{XUV}}/4\pi a^2 G m_{\text{core}}$ where $\eta_p, r_{\text{core}}, a, m_{\text{core}}$ are the planet's
mass-loss efficiency, core radius, orbital separation, and core mass
respectively, $L_{\text{XUV}}$ is the XUV luminosity of the host star and $G$
is the gravitational constant. By writing the atmospheric mass as the product
of the planet mass and envelope mass fraction ($M_{\text{atm}}=m_p\: X_2$), the
mass loss timescale under photoevaporation
($t_{\text{loss}} = M_{\text{atm}}/ \dot{M}_{\text{atm}}$) scales as

\begin{align}
t_{\text{loss}} &\propto \frac{m_p^2\: a^2\: X_2}{\eta_p\: r_{\text{core}}^{3}\: L_{\text{XUV}}} \nonumber \\
&\propto \frac{m_p^3\: a^2\: X_2}{r_{\text{core}}^{4}\: L_{\text{XUV}}}
\label{eq:tlpe}
\end{align}

\noindent where we have adopted
$\eta_p \propto v_{\text{esc}}^{-2} \propto m_{\text{core}}^{-1} r_{\text{core}}$ 
\citep{owen17} and set $m_{\text{core}}=m_p$ by assuming that the planet
masses are dominated by their rocky core masses. In this simple picture,
\cite{owen20} set \autoref{eq:tlpe} to the maximum mass loss timescale for a
rocky planet below the valley which is assumed to have just lost the entirety
of its initial H/He envelope. In order to form the radius valley, this
timescale must be less than the maximum timescale for 
the gaseous (i.e. non-rocky) planet to have retained its initial H/He envelope
with an atmospheric mass fraction of $X_2$. This criterion places the
following constrains on the rocky and gaseous planet parameters according to

\begin{align}
\frac{t_{\text{loss,gas}}}{t_{\text{loss,rock}}} \geq 1, \nonumber \\
\left( \frac{m_{p,\text{gas}}}{m_{p,\text{rock}}} \right)
\left( \frac{a_{\text{gas}}}{a_{\text{rock}}} \right)^{2/3}
\left( \frac{r_{\text{core,gas}}}{r_{\text{core,rock}}} \right)^{-4/3} \geq 1.
\label{eq:pev2}
\end{align}

\noindent The power of comparing planets within the same planetary system
is evidenced in \autoref{eq:pev2} in which the unknown quantity $L_{\text{XUV}}$
is scaled out of the expression.

In the photoevaporation model, the stripped rocky planet has been reduced to
its Earth-like core such that the core radius is equivalent to the planet's
radius; $r_{\text{core,rock}} = r_{p,\text{rock}}$. Noting
that $r_{\text{core}} \propto m_{\text{core}}^{0.27}$ for Earth-like
bodies \citep{zeng16}, we write $r_{\text{core,gas}} = m_{p,gas}^{0.27}$ where the
input radius and mass are each given in units of the Earth. It follows from
\autoref{eq:pev2} that the minimum mass of the gaseous planet under the
photoevaporation model is

\begin{equation}
\frac{m_{p,\text{gas}}}{\text{M}_{\oplus}} \geq 
\left[ \left( \frac{m_{p,\text{rock}}}{\text{M}_{\oplus}} \right)
\left( \frac{a_{\text{rock}}}{a_{\text{gas}}} \right)^{2/3}\:
\left( \frac{r_{p,\text{rock}}}{\text{R}_{\oplus}} \right)^{-4/3} \right]^{1.56}.
\label{eq:pev3}
\end{equation}

\noindent The inequality in \autoref{eq:pev3} must be satisfied for the
planetary parameters to be consistent with the photoevaporation model.
Similarly,

\begin{equation}
\frac{m_{p,\text{rock}}}{\text{M}_{\oplus}} \leq 
\left( \frac{m_{p,\text{gas}}}{\text{M}_{\oplus}} \right)^{0.64}
\left( \frac{a_{\text{gas}}}{a_{\text{rock}}} \right)^{2/3}\:
\left( \frac{r_{p,\text{rock}}}{\text{R}_{\oplus}} \right)^{4/3}
\label{eq:pev4}
\end{equation}

\noindent represents the maximum mass of the rocky planet for the system to
be consistent with the photoevaporation model.

A few notable caveats exist with this simplified model \citep{owen20}.
Specifically, these calculations assumed
that the envelope mass fraction $X_2$ for which the mass loss timescales are
maximized, is independent of the planet properties. Furthermore, individual
gaseous planets may have envelope mass fractions that are greater than that
which is required to maximize $t_{\text{loss,gas}}$. Lastly, this simplified model
ignores the contraction of the H/He envelope over time. This poses a critical
limitation as gaseous envelopes are likely to have been more extended at
early times when photoevaporation was actively ongoing, compared to their
present day values.

These issues are alleviated by the \texttt{EvapMass} software \cite{owen20}
which calculates the value
of $X_2$ that maximizes the mass loss timescale and self-consistently models
the gaseous envelope structure from the typical Kelvin-Helmholtz time of the
gaseous
envelope $(\tau_{\text{KH}}\sim 100$ Myrs) to the present. However, attempting
these numerical calculations on the LTT 3780 system resulted in a failure to
solve for a lower limit on the LTT 3780c core mass. By default,
\texttt{EvapMass} only
considers $m_{\text{core,gas}}\geq 0.1$ \Mearth{,} which is itself a very weak
constraint, such that the \texttt{EvapMass}
calculation does not provide any new insight into the minimum mass of LTT 3780c.

\section{Limits on the planet masses for consistency with models of
core-powered mass loss} \label{app:cpml}
Here we derive the constraints on the planet masses in order to be consistent
with the core-powered mass loss model for sculpting the radius valley. 
Analogously to the formalism presented in \autoref{app:pe}, we
compare the timescales for core-powered mass loss of planets spanning the
radius valley and within the same multi-transiting system.

Core-powered mass loss
is another mechanism for driving thermal escape of a planet's atmosphere due
to the planetary core's own cooling luminosity \citep{ginzburg18,gupta19}.
Similarly to the initial conditions assumed in the photoevaporation model,
here a population of non-rocky planets is formed with a distribution of
Earth-like core masses plus H/He envelopes. Their atmospheres are described by a
lower convective region which is terminated at the radius of the
radiative-convective boundary (RCB), above which the atmosphere becomes
isothermal and heat is transported radiatively to the planet's
Bondi radius. The Bondi radius is set by equating the planet's escape velocity
to its thermal sound speed and is $R_{\text{B}} = Gm_{\text{core}}/c_s^2$, where
$G$ is the gravitational constant, $m_{\text{core}}$ is the core mass, and
the thermal sound speed is $c_s = \sqrt{k_B T_{\text{eq}}/\mu}$, where
$k_B$ is the Boltzmann constant, \teq{} is the equilibrium temperature,
and $\mu$ is the atmospheric mean molecular
weight, assumed to be 2 amu for H$_2$. The Bondi-limited regime
represents the physical limit of the atmospheric mass loss rate and is dictated
by the gas thermal velocity at $R_{\text{B}}$.

The corresponding Bondi-limited mass loss rate is
$\dot{M}_{\text{atm}}=4\pi R_{\text{B}}^2 c_s \rho_{\text{RCB}} \exp{(-G m_{\text{core}}/c_s^2 R_{\text{RCB}})}$ where $\rho_{\text{RCB}}$ is the atmospheric density at
the RCB whose radius is
$R_{\text{RCB}}$. The majority of the atmosphere's mass lies within its
convective zone such that integrating an adiabatic gas density profile over the
convective zone returns the approximate atmospheric mass 

\begin{equation}
M_{\text{atm}} \approx 4\pi R_{\text{RCB}}^3 \rho_{\text{RCB}}
\left( \frac{\gamma-1}{\gamma}
\frac{R_{\text{B}}}{R_{\text{RCB}}} \right)^{1/(\gamma-1)}
\end{equation}

\noindent where $\gamma$ is the adiabatic index which is fixed to $4/3$
\citep{ginzburg16}. The resulting mass loss timescale
($t_{\text{loss}} = M_{\text{atm}}/ \dot{M}_{\text{atm}}$) scales as

\begin{align}
t_{\text{loss}} &\propto \frac{R_{\text{B}}}{c_s}\:
\exp{\left( \frac{G m_{\text{core}}}{c_s^2 R_{\text{RCB}}} \right)}, \nonumber \\
&\propto \frac{m_p}{T_{\text{eq}}^{3/2}}\:
\exp{\left( \frac{c' m_p}{T_{\text{eq}} r_p} \right)}, \\
\end{align}

\noindent where the constant
$c' =G\mu/k_B\sim 10^{4}$ \Rearth{} K \Mearth{$^{-1}$}, the planet's envelope
mass fraction is assumed to be small such that $m_{\text{core}}\approx m_p$, and
the $R_{\text{RCB}}$ is treated as the planet's effective radius;
$R_{\text{RCB}} \approx r_p$.

Analogously to the photoevaporation scenario,
for the planetary parameters within a multi-transiting system and spanning the
radius valley to be consistent with the core-powered mass loss scenario,  
we require that the mass loss timescale of the gaseous (i.e. non-rocky) planet
exceeds that of the rocky planet. This leads to the following condition:

\begin{align}
\frac{t_{\text{loss,gas}}}{t_{\text{loss,rock}}} &\geq 1, \nonumber \\
\left( \frac{m_{p,\text{gas}}}{m_{p,\text{rock}}} \right)
\left( \frac{T_{\text{eq,gas}}}{T_{\text{eq,rock}}} \right)^{-3/2} \nonumber \\
\exp{\left[ c' \left( \frac{m_{p,\text{gas}}}{T_{\text{eq,gas}}\: r_{p,\text{gas}}} - \frac{m_{p,\text{rock}}}{T_{\text{eq,rock}}\: r_{p,\text{rock}}} \right) \right]} &\geq 1.
\label{eq:cpmlv2}
\end{align}

\noindent The appearance of the planet masses as both 
linear factors and in the exponential function means that \autoref{eq:cpmlv2}
belongs to the class of Lambert $W$ functions of the form
$f(m)\propto m\text{e}^m$. Such functions
do not have analytical solutions but the limiting planet masses under the
core-powered mass loss model can be solved for numerically.